\documentclass[aps,pre,twocolumn,superscriptaddress]{revtex4-1}
\usepackage{float}
\usepackage{color}
\usepackage{graphicx}
\usepackage{epstopdf}
\begin{document}

% Use the \preprint command to place your local institutional report
% number in the upper righthand corner of the title page in preprint mode.
% Multiple \preprint commands are allowed.
% Use the 'preprintnumbers' class option to override journal defaults
% to display numbers if necessary
%\preprint{}

%Title of paper
\title{Double Inverse Stochastic Resonance with Dynamic Synapses}

% repeat the \author .. \affiliation  etc. as needed
% \email, \thanks, \homepage, \altaffiliation all apply to the current
% author. Explanatory text should go in the []'s, actual e-mail
% address or url should go in the {}'s for \email and \homepage.
% Please use the appropriate macro foreach each type of information

% \affiliation command applies to all authors since the last
% \affiliation command. The \affiliation command should follow the
% other information
% \affiliation can be followed by \email, \homepage, \thanks as well.
%
%\author{}
%\email[]{Your e-mail address}
%\homepage[]{Your web page}
%\thanks{}
%\altaffiliation{}
%\affiliation{}

\author{Muhammet Uzuntarla}
\email[]{muzuntarla@yahoo.com}
%\homepage[]{Your web page}
%\thanks{}
%\altaffiliation{}
\affiliation{Department of Biomedical Engineering, Bulent Ecevit University, Zonguldak, Turkey}

\author{Joaquin J. Torres}
%\email[]{Your e-mail address}
%\homepage[]{Your web page}
%\thanks{}
%\altaffiliation{}
\affiliation{Department of Electromagnetism and Physics of the Matter and Institute Carlos I for Theoretical and Computational Physics, University of Granada, Granada, E-18071 Spain}

\author{Paul So}
%\email[]{Your e-mail address}
%\homepage[]{Your web page}
%\thanks{}
%\altaffiliation{}
\affiliation{Department of Physics and Astronomy and The Krasnow Institute for Advanced Study, George Mason University, Fairfax, Virginia, USA}

\author{Mahmut Ozer}
%\email[]{muzuntarla@yahoo.com}
%\homepage[]{Your web page}
%\thanks{}
%\altaffiliation{}
\affiliation{Department of Electrical and Electronics Engineering, Bulent Ecevit University, Zonguldak, Turkey}

\author{Ernest Barreto}
%\email[]{Your e-mail address}
%\homepage[]{Your web page}
%\thanks{}
%\altaffiliation{}
\affiliation{Department of Physics and Astronomy and The Krasnow Institute for Advanced Study, George Mason University, Fairfax, Virginia, USA}

%Collaboration name if desired (requires use of superscriptaddress
%option in \documentclass). \noaffiliation is required (may also be
%used with the \author command).
%\collaboration can be followed by \email, \homepage, \thanks as well.
%\collaboration{}
%\noaffiliation

\date{\today}

\begin{abstract}
We investigate the behavior of a model neuron that receives a biophysically-realistic noisy post-synaptic current based on uncorrelated spiking activity from a large number of afferents. We show that, with static synapses, such noise can give rise to inverse stochastic resonance (ISR) as a function of the presynaptic firing rate. We compare this to the case with dynamic synapses that feature short-term synaptic plasticity, and show that the interval of presynaptic firing rate over which ISR exists can be extended or diminished. We consider both short-term depression and facilitation. Interestingly, we find that a double inverse stochastic resonance (DISR), with two distinct wells centered at different presynaptic firing rates, can appear.
\end{abstract}

% insert suggested PACS numbers in braces on next line

% insert suggested keywords - APS authors don't need to do this
%\keywords{}

%\maketitle must follow title, authors, abstract, \pacs, and \keywords
\maketitle

\section{Introduction}

It is widely accepted that noise can facilitate the information processing capabilities of neurons \cite{McDonell2011,lindner04,collins1996,bezrukov95,Ozer2009964}. A well-known example is stochastic resonance (SR), where a certain amount of noise can enhance the detection of weak signals in a neural medium \cite{benzi1981,collins_stochastic_1995,gammaitoni98,manjarrez02,Ozer2009964}.
More precisely, for low noise levels the system is not able to detect a weak signal due to its small amplitude. For moderate noise levels, however, the noise enhances the signal up to a detection
threshold. This makes the neurons respond in a manner that is strongly correlated with the signal. Conversely, for strong noise, neural activity is highly
variable and little correlated with the signal. This behavior of the neural response results in a bell-shaped dependence on noise, indicating that the correlation between the signal and neural activity is maximal around a moderate level of noise. 

On the other hand, the repetitive spiking behavior of a single neuron driven just above its spiking threshold can be inhibited by the presence of noise, as was demonstrated by a series of theoretical studies \cite{TuckwellPhysRevE09,Guo11,ISRtwoneurons,Uzuntarlasolo13,uzuntarla13,schmerl13}. In these works, the existence of a pronounced minimum in the average firing rate as a function of  the noise level was reported. This noise-driven inhibitory effect has also been shown experimentally in \emph{in vitro} preparations of squid axons which operate as pacemakers \cite{paydarfar_noisy_2006}. Since the dependence of the neuronal response on noise is the opposite of that in the SR mechanism, this phenomenon is called inverse stochastic resonance (ISR).

It is widely assumed that neurons transmit information and communicate with each other through spikes. Thus, the emergence of ISR can be seen as a limitation for information processing
in neural systems. However, ISR might play an important role in computational mechanisms that require reduced firing activity without chemical inhibitory neuromodulation, or alternatively, when other computational mechanisms require on-off bursts of activity \cite{schmerl13}. In this sense, ISR may be advantageous for such particular cases.

In recent years, there has been a growing interest in studying different aspects of the ISR phenomenon. For instance, Guo \cite{Guo11} investigated the influence of temporal noise correlations on ISR, showing that colored noise suppresses neural activity more strongly than white (Gaussian) noise. Tuckwell and Jost \cite{tuckwell_effects_2011} considered ISR in a more realistic neuron model with spatial extent. They showed that if the signal and noise inputs overlap spatially on the neuron, then weak noise may inhibit spiking. If, however, the signal and noise are non-uniformly applied, then the noise has no effect on the spiking activity, no matter how large its region of application is. 

Previous works on ISR focused on external rather than endogenous noise sources \cite{TuckwellPhysRevE09,Guo11, ISRtwoneurons}. However, insofar as in vivo neural activity is inherently noisy \cite{Stein2005, Lin2015644}, those works failed to account for the actual biophysical conditions and did not provide a clear understanding of the phenomenon under realistic conditions. In this context, recent studies have examined the possible biological mechanisms that might give rise to ISR in actual neural systems. For instance, in \cite{uzuntarla13}, noise was considered to arise from the stochastic nature of voltage gated ion channels embedded in neural membranes. Under this biophysically realistic scenario, the authors showed that ISR can indeed arise in a single neuron due to ion channel noise, where its strength is proportional to cell membrane area (see also \cite{schmerl13}). In addition, the authors clarified the dynamical structure underlying the ISR phenomenon. In another recent work \cite{Uzuntarlasolo13}, ISR was studied using a detailed modeling approach for the synaptic background activity by assuming that spike transmission from presynaptic afferents to the postsynaptic neuron is unreliable, and it was shown that unreliable synaptic transmission might be a potential biophysical mechanism that gives rise to ISR.

In the present study, we continue to investigate the ISR phenomenon in a single neuron under more realistic conditions by considering the underlying noise as originating from the presynaptic background activity. Stochastic neurotransmitter release, varying quantities of neurotransmitter resources at different synapses, and spatial heterogeneity of synaptic transmission along the dendritic tree \cite{fssjn03} can all modulate the neuron's postsynaptic response. Collectively, these processes contribute to short- term synaptic plasticity, that is the transient change of synaptic efficiency on time scales from millisecond to minutes. Short-term plasticity may come in different flavors, i.e., short-term depression (STD) and/or facilitation (STF). From a mechanistic point of view in particular, STD refers to the progressive reduction of synaptic neurotransmitter release by depletion of available neurotransmitter resources at the synapse \cite{zuckerARP02,dittmanJNEUROSCI,Cho13042011,delarocha05}. STF refers instead to transient increases of neurotransmitter release probability by activity-dependent presynaptic accumulation of intracellular {{Ca$^{2+}$}} \cite{bertramJNEURO,dittmanJNEUROSCI,synapticnoise,Cho13042011}.

Since these two synaptic mechanisms have been widely reported to be relevant for dynamics of neural circuits \cite{tsodyksNC,Senn98,Dror2000,tsodyksjn00,tsodyksCODING,torresNC,tsodyks06,barak2007,torresNC2007,mejiasCD08,mejias2011emergence,TorresKappen2013} and various brain functions \cite{torresCAPACITY,MacLeod2863,Fortune2002539,mejias09,mejiasupdown10,Bourjaily513,mejias12}, here we investigate how short-term synaptic plasticity can influence the ISR phenomenon. {{With this aim we analyze}} the response of a postsynaptic neuron that receives presynaptic inputs from a set of afferents through dynamic synapses with short-term synaptic plasticity and examine the influence of various parameters of the dynamic synapse mechanisms on different ISR features. We show that when both STD and STF mechanisms are included, a new intriguing phenomenon may be observed consisting of two ISRs occurring at distinct presynaptic firing rates. {{We call this phenomenon double}} inverse stochastic resonance (DISR). We conclude with a discussion of the possible computational implications of our findings for an actual neural system. 

\section{Models and Methods}

The system under study consists of a postsynaptic neuron that receives uncorrelated network activity from a finite number of excitatory and inhibitory presynaptic neurons through dynamical synapses. The temporal dynamics of the postsynaptic membrane potential is modeled according to Hodgkin and Huxley \cite{HHb} and reads
\begin{eqnarray}
C\frac{dV}{dt} = && I_{0 }-g_{Na}m^{3}h\left(V-E_{Na}\right) -g_{K}n^{4}\left(V-E_{K}\right) \nonumber \\
&& -g_{L}\left(V-E_{L}\right) +I_{syn}
\end{eqnarray}
where $V$ denotes the membrane potential in millivolts, and $C$ is the membrane capacitance per unit of membrane area. $I_{0}$ is an injected external bias current density, and is used for the modulation of neuronal excitability. Unless stated otherwise, we set $I_{0}$ to a value (see Table I) for which the neuron exhibits bistability between a silent (equilibrium) and a spiking (limit cycle) state \cite{TuckwellPhysRevE09,Guo11,uzuntarla13}. $g_{Na}$, $g_{K}$, and $g_{L}$ are the maximum conductances for sodium, potassium, and leak channels, respectively. $E_{Na}$, $E_{K}$, and $E_{L}$ denote the corresponding sodium, potassium, and leak reversal potentials. The values used for all parameters, unless otherwise noted, are listed in Table I.

\begin{table}[ht]
\caption{Neuron model parameters}
\centering
\begin{tabular}{l l l l}
\hline\hline
Symbol$\quad$& Description & Value & Units \\ [0.5ex] % inserts table %heading
\hline
$C$$\quad$ & Membrane capacitance$\quad$ & $1$ & $\mathrm{\mu F/cm^{2}}$ \\ [0.5ex] % inserts table %heading
$g_{Na}\quad$ & Maximum sodium conductance$\quad$ & $120$ & $\mathrm{mS/cm^{2}}$ \\ [0.5ex] % inserts table %heading
$g_{K}\quad$ & Maximum potasium conductance$\quad$ & $36$ & $\mathrm{mS/cm^{2}}$ \\ [0.5ex] % inserts table %heading
$g_{L}\quad$ & Leakage conductance$\quad$ & $0.3$ & $\mathrm{mS/cm^{2}}$ \\ [0.5ex] % inserts table %heading
$E_{Na}\quad$ & Sodium reversal potential$\quad$& $115$ & mV \\ [0.5ex] % inserts table %heading
$E_{K}\quad$ & Potassium reversal potential$\quad$ & $-12$ & mV \\ [0.5ex] % inserts table %heading
$E_{L}\quad$ & Leakage reversal potential$\quad$ & $10.6$ & mV \\ [0.5ex] % inserts table %heading
$I_{0}\quad$ &  Injected bias current $\quad$ & $6.8$ & $\mathrm{\mu A/cm^{2}}$ \\ [0.5ex] % inserts table %heading

\hline
\end{tabular}
\label{table:nonlin}
\end{table}

 The gating variables $m$, $h$, and $n$ model the activation and inactivation of the sodium channels and the activation of the potassium channels, respectively, and obey the following differential equations \cite{HHb}
\begin{eqnarray}
\frac{dm}{dt} &=& \alpha_m(V) (1-m) - \beta_m(V)m \nonumber \\
\frac{dn}{dt} &=& \alpha_n(V) (1-n) - \beta_n(V)n  \\
\frac{dh}{dt} &=& \alpha_h(V) (1-h) - \beta_h(V)h\nonumber
\end{eqnarray} where $\alpha_{\rho}$ and $\beta_{\rho}$ $(\rho=m,\, n,\, h)$ are experimentally-determined voltage-dependent rate functions, defined by \cite{HHb, pankratova2005b}:

\begin{eqnarray}
\alpha_m(V) &=& 0.1\frac{(25-V)}{\exp[(25-V)/10]-1}  \nonumber \\
\beta_m(V) &=& 4\exp[-V/10]  \nonumber \\
\alpha_n(V) &=& 0.01\frac{(10-V)}{\exp[(10-V)/10]-1}  \\
\beta_n(V) &=& 0.125\exp[-V/80]  \nonumber \\
\alpha_h(V) &=& 0.07\exp[-V/20] \nonumber \\
\beta_h(V) &=& \frac{1}{\exp[(30-V)/10]+1}.  \nonumber
 \end{eqnarray}

In our setup, presynaptic neurons are modeled as independent Poisson spike generators emitting uncorrelated spikes with frequency $f$. For synaptic transmission to the postsynaptic neuron, we adopt the dynamic synapse formulation of Tsodyks and Markram \cite{tsodyksNC}. This model considers the total amount of neurotransmitter resources at synapse $i$ to be proportioned among three states: ``available'' $x_i(t)$, ``active'' $y_i(t)$, and ``inactive'' $z_i(t)$, normalized such that $x_i+y_i+z_i=1$. When a spike arrives at synapse $i$ at time $t$, there is an instantaneous transfer of a fraction $u_{i}(t) \in [0,1]$ of available resources $x_i(t)$ to the active state. Active resources then deactivate over a time on the order of a few milliseconds, characterized by the time constant $\tau_{in}$, and recover over a time period characterized by $\tau_{\ensuremath{rec}}$, which can vary from tens of milliseconds to seconds depending on the type of neuron and the type of synapse. The dynamical behavior of these synaptic states is governed by a system of three coupled differential equations as follows \cite{tsodyksNC}: 

\begin{equation}
\frac{dx_{i}(t)}{dt}=\frac{z_{i}(t)}{\tau_{rec}}-u_{i}(t)x_{i}(t)\delta(t-t_{spike}^{i})\;\;
\end{equation}
\begin{equation}
\frac{dy_{i}(t)}{dt}=-\frac{y_{i}(t)}{\tau_{in}}+u_{i}(t)x_{i}(t)\delta(t-t_{spike}^{i})
\end{equation}

\begin{equation}
\frac{dz_{i}(t)}{dt}=\frac{y_{i}(t)}{\tau_{in}}-\frac{z_{i}(t)}{\tau_{rec}}\quad\quad\quad\quad\quad\quad\quad
\end{equation}
where the delta function refers to the arrival time of a spike at synapse $i$ at $t=t_{spike}^{i}$. The model described by Eqs.~$(3-5)$ reproduces short-term depression (STD) phenomena in cortical neurons for relatively long $\tau_{rec}$, assuming a constant transmitter release fraction $u_{i}(t)=\mathcal{U}$ \cite{tsodyksPNAS,tsodyksNC}.

To include short-term facilitation (STF) in the model, the release fraction $u_{i}(t)$ is allowed to vary in time. The equation below models the dependence of $u_{i}(t)$ on the intracellular calcium concentration, which increases due to calcium influx through voltage-sensitive calcium channels after the arrival of successive spikes \cite{bertramJNEURO}:
\begin{equation}
\frac{du_{i}(t)}{dt}=\frac{\mathcal{U}-u_{i}(t)}{\tau_{fac}}+\mathcal{U}[1-u_{i}(t)]\delta(t-t_{spike}^{i}).
\label{dynueq}
\end{equation}
Here, $\mathcal{U}$ is the release fraction at rest, and the calcium dynamics enters through $\tau_{\ensuremath{fac}}$, the characteristic time for the calcium channel gates to transition from the open to the closed state (which terminates the calcium influx) \cite{tsodyksNC}. In this model, the level of STD and STF at a synapse can be controlled by the corresponding time constants, so that larger values of $\tau_{\ensuremath{rec}}$ and $\tau_{\ensuremath{fac}}$ {{respectively result in}} stronger synaptic depression and facilitation effects. For facilitating synapses, a larger value of $\mathcal{U}$ produces a larger increase in the release fraction after each spike, thus inducing stronger and faster facilitation. However, note that larger values of $\mathcal{U}$ in a depressing synapse mean that more synaptic resources are used per spike, which can lead to faster depletion and more depression for subsequent spikes, particularly at high presynaptic firing rates.

In Eq.~(1), $I_{syn}$ represents the total synaptic current generated by $N=N_e+N_i=1000$ presynaptic excitatory and inhibitory inputs, where the ratio  $N_{e}:N_{i}$ is set to $4:1$,
{{as observed \emph{in vivo} conditions}} \cite{brunel00}.  Then, based on the synaptic dynamics described above, the postsynaptic current generated at synapse $i$ is taken to be proportional to the amount of active neurotransmitter, namely $I_{i}(t)=\mathcal{A}y_{i}(t).$ Here, $\mathcal{A}$ is the maximum postsynaptic current which can be generated at the synapse by activating all resources. Accordingly, the total postsynaptic current introduced into the postsynaptic neuron due to the arrival of both excitatory and inhibitory presynaptic inputs is

\begin{equation}
I_{syn}(t)=\sum_{p=1}^{N_{e}}\mathcal{A}y_{p}(t)-K\sum_{q=1}^{N_{i}}\mathcal{A}y_{q}(t)
\end{equation}
where $K$ is the relative strength between inhibitory and excitatory connections. We set $K=4$ so that the average postsynaptic current is zero, corresponding to the physiological range of balanced states of cortical neurons \cite{Braitenberg91}. Such a state allows us to control the excitability of the postsynaptic neuron using only $I_{0}$. We denote the standard deviation of $I_{syn}$ by $\sigma_I$.

To characterize ISR quantitively, we follow the procedures used in \cite{gutkin_inhibition_2009,uzuntarla13,TuckwellPhysRevE09}. For trial $j$ $($j=1$ \dots $L$)$, an initial condition is randomly selected for the postsynaptic neuron with uniform probability within a fixed region of the four-dimensional state space ($V,\, m,\, n$, and $h).$ Specifically, this region ranges from $-10$ to $80\, \mathrm{mV}$ for the membrane voltage variable $V$, and from $0$ to $1$ for the each of the gating variables $m,\, n$, and $h$. Then, the system equations are integrated for a transient time $T=1\,\mathrm{s}$. After this, we count the number of spikes generated by the postsynaptic neuron $N_{spikes}$ that occur in an additional time interval of duration $\Delta t=5\, \mathrm{s}$. This entire procedure is repeated $L$ times, and the mean firing rate is calculated as follows: 

\begin{center}
\begin{equation}
\nu=\frac{1}{L\Delta t}\left(\sum\limits _{j=1}^{L}N_{spikes}^{j}\right)
\end{equation}

\par\end{center}

\noindent The results presented in the next sections are obtained
over $L=1000$ independent runs for each set of parameter values to
warrant appropriate statistical accuracy with respect to the stochastic
fluctuations in background activity.
\noindent\begin{figure*}[ht!]
\begin{centering}
\includegraphics[height=5cm, width=8cm]{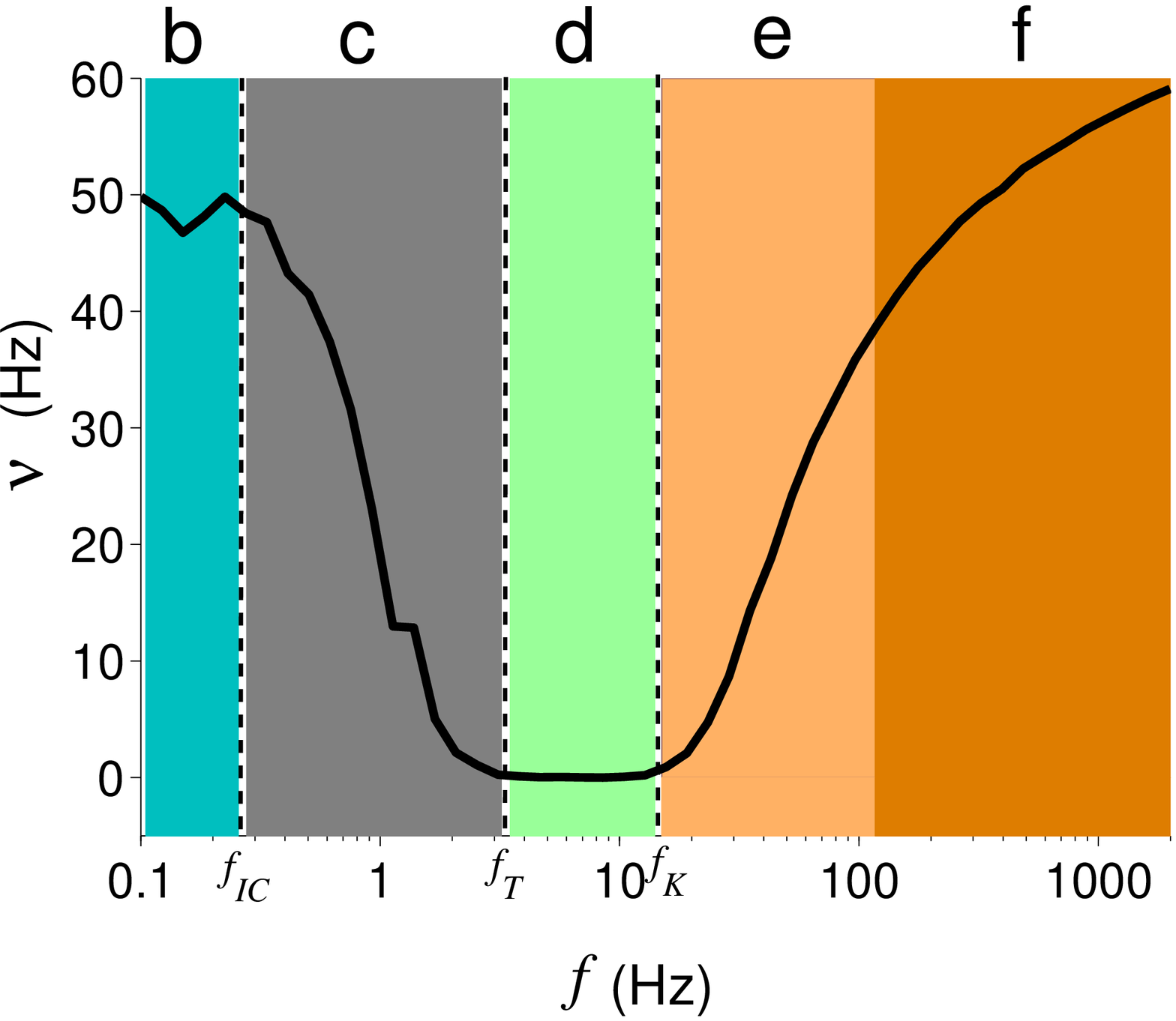}~\includegraphics[ height=5cm, width=8cm]{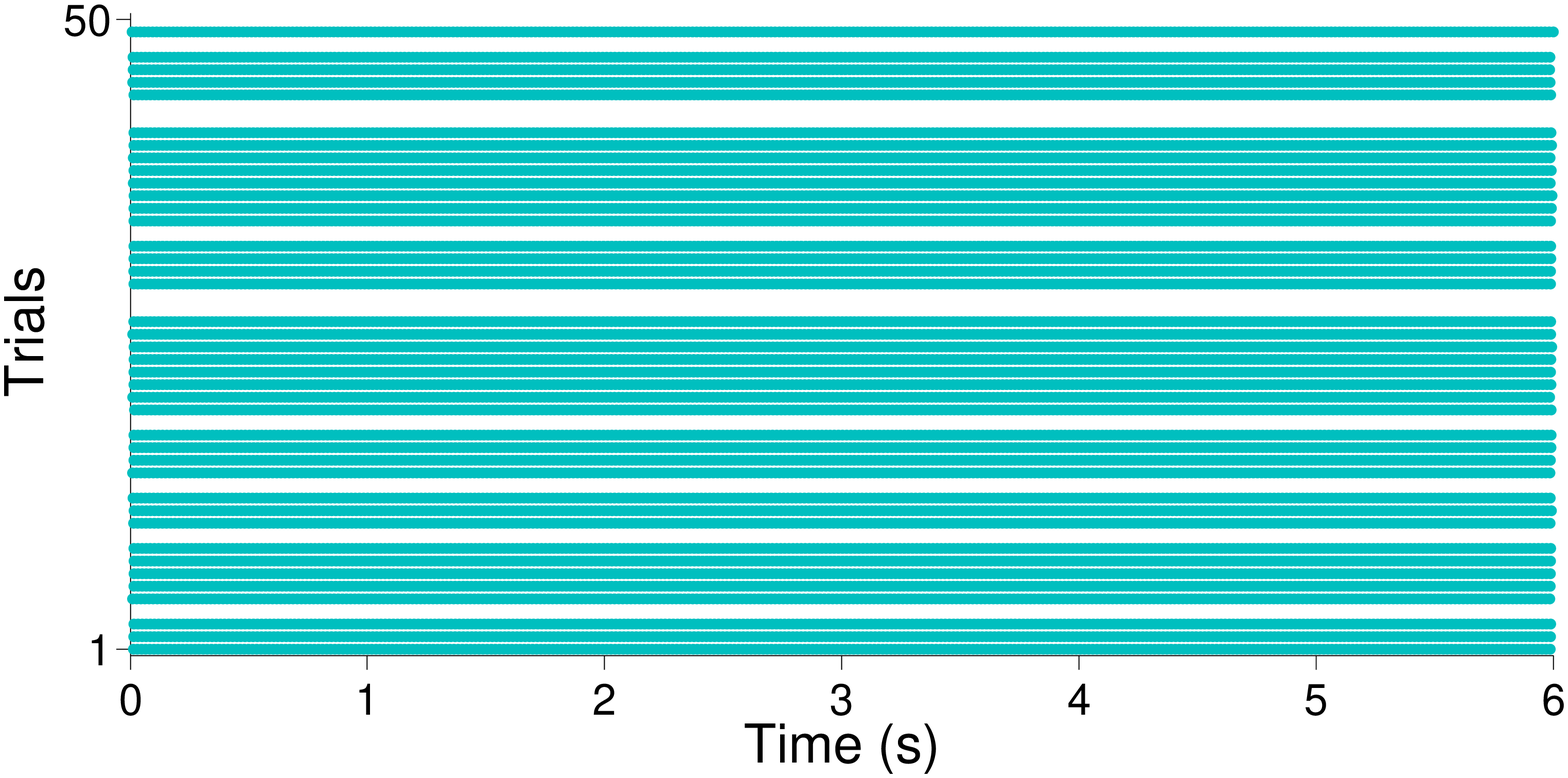}
\par\end{centering}
\begin{centering}
(a)~~~~~~~~~~~~~~~~~~~~~~~~~~~~~~~~~~~~~~~~~~~~~~~~~~~~~~~~~~~~~~~~~~~~~~~~~
(b)
\par\end{centering}
\begin{centering}
\includegraphics[ height=5cm, width=8cm]{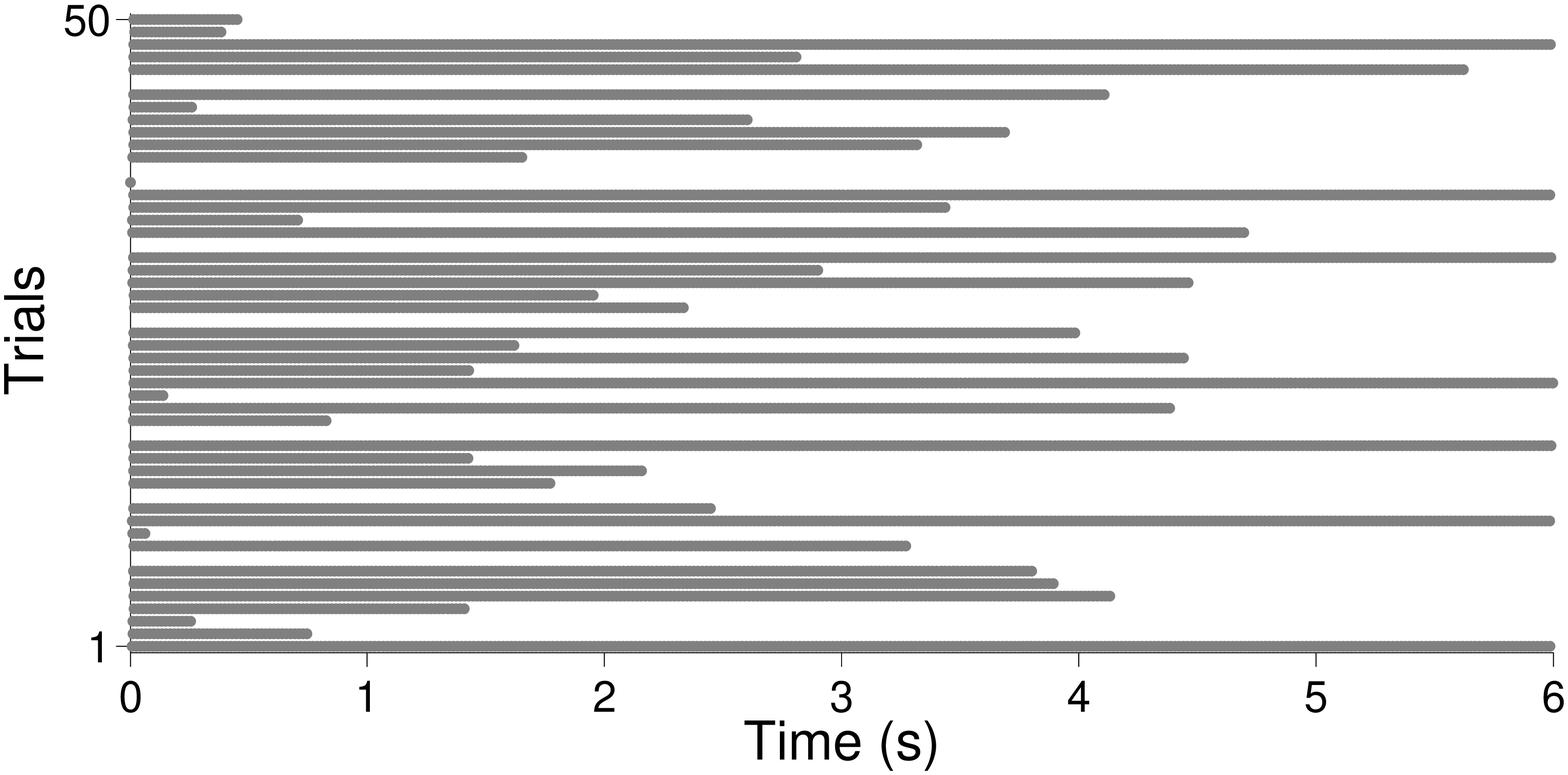}\includegraphics[ height=5cm, width=8cm]{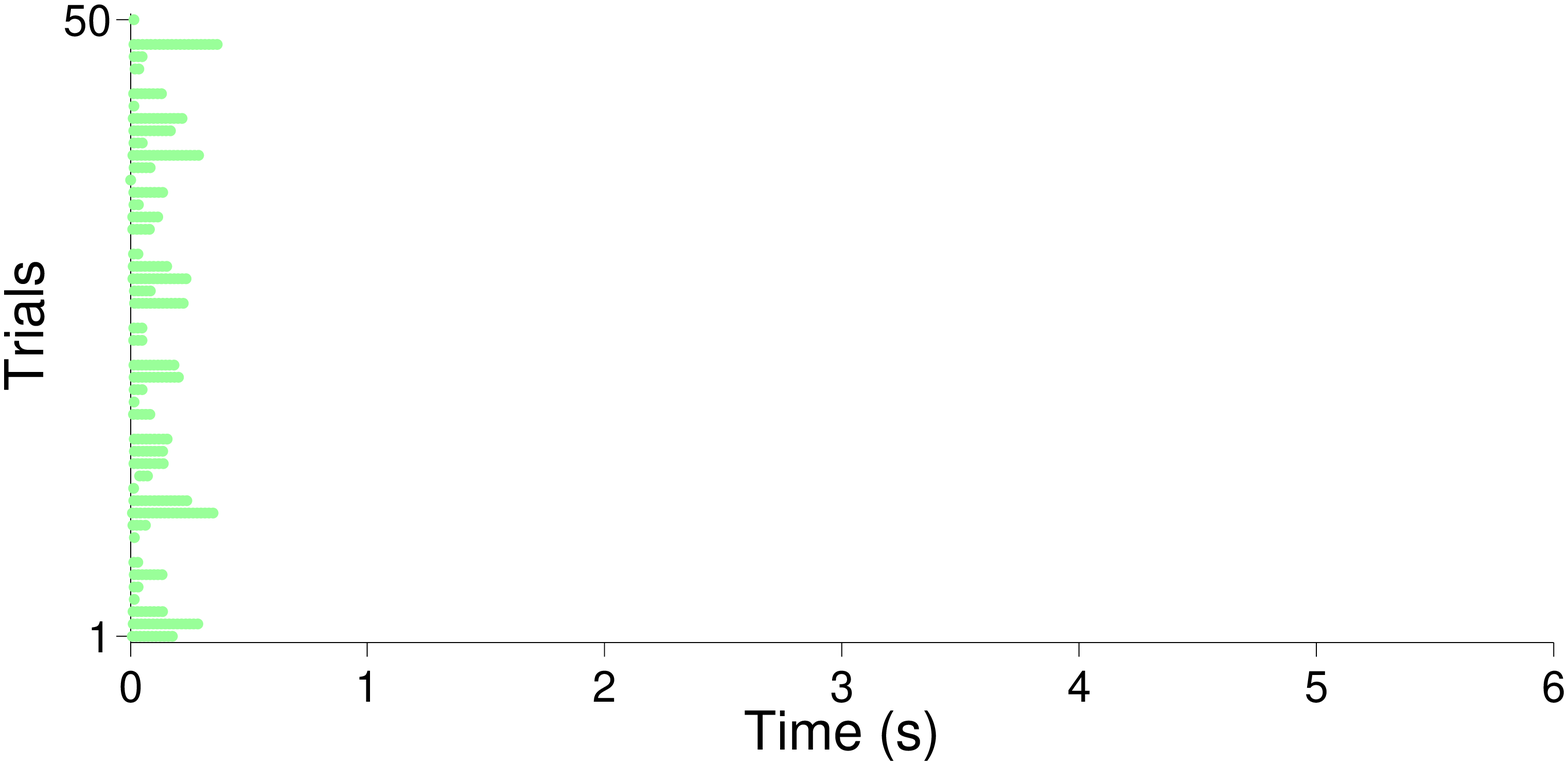}
\par\end{centering}
\begin{centering}
(c)~~~~~~~~~~~~~~~~~~~~~~~~~~~~~~~~~~~~~~~~~~~~~~~~~~~~~~~~~~~~~~~~~~~~~~~~~
(d)
\par\end{centering}
\begin{centering}
\includegraphics[ height=5cm, width=8cm]{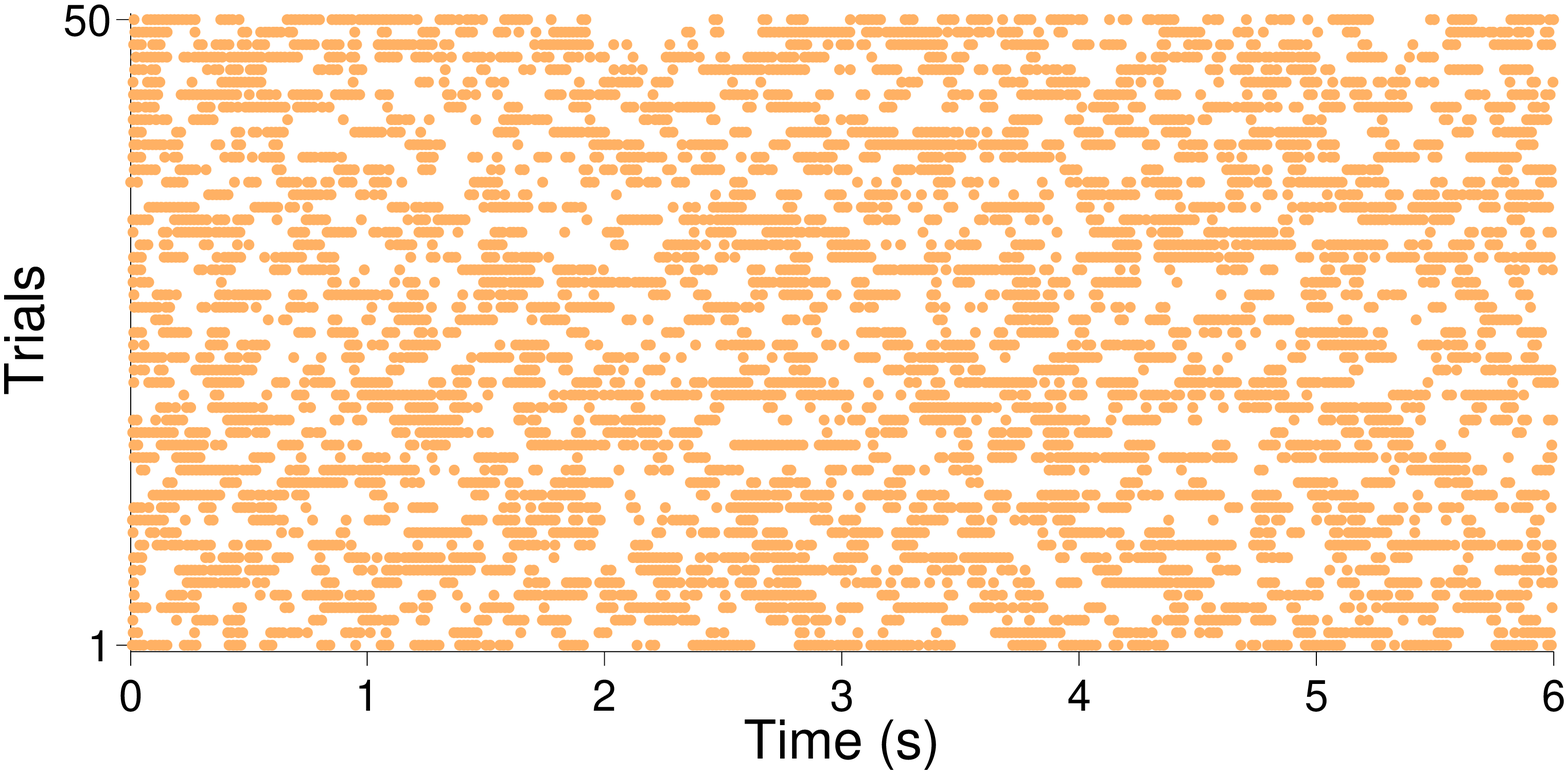}\includegraphics[ height=5cm, width=8cm]{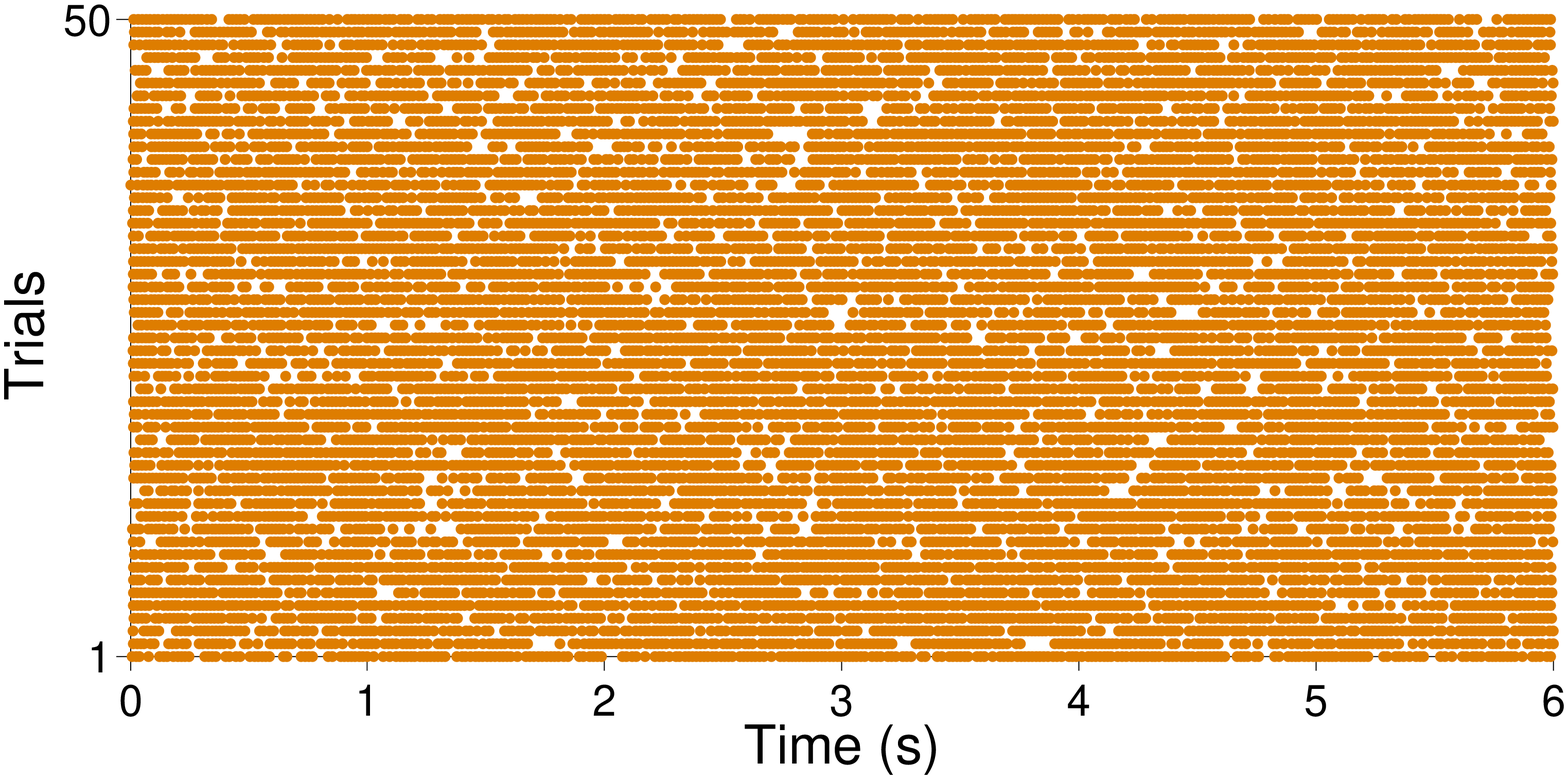}
\par\end{centering}
\begin{centering}
(e)~~~~~~~~~~~~~~~~~~~~~~~~~~~~~~~~~~~~~~~~~~~~~~~~~~~~~~~~~~~~~~~~~~~~~~~~~
(f)
\par\end{centering}
\caption{\label{fig1} \textbf{(a)} (Color online) ISR in the response of the postsynaptic neuron with static synapses. The mean firing rate $\nu$ is plotted versus the presynaptic firing rate $f$. Vertical dashed lines correspond to presynaptic firing frequencies where the firing behavior of the postsynaptic neuron exhibits transitions as follows: For $f<f_{IC}$, only the initial condition effect exists {{(region ``b'')}}. For $f_{IC}<f<f_T$, the trapping effect begins to appear {{(region ``c'')}}. For $f_{T}<f<f_K$, trapping occurs immediately and in all trials {{(region ``d'')}}. For $f>f_{K}$, the kickout effect increasingly dominates {{(regions ``e'' and ``f'')}}. \textbf{(b)-(f)} Sample raster plots of postsynaptic neuron spiking activity across $50$ trials for different regions (marked with the corresponding letters and colors) of the ISR curve in panel a. The panels are constructed by plotting a dot to indicate the occurrence of a postsynaptic spike at a given time in a given trial. The mean frequency of the Poissonian presynaptic spike trains that generate the background activity is \textbf{(b)} $f=0.1\,\mathrm{Hz}$, \textbf{(c)} $f=1\,\mathrm{Hz}$, \textbf{(d)} $f=10\,\mathrm{Hz}$, \textbf{(e)} $f=50\, \mathrm{Hz}$, and \textbf{(f)} $f=200\, \mathrm{Hz}$.}
\end{figure*}\\

\section{Results}

\subsection{ISR with static synapses}
As a first step towards understanding the implications of short-term synaptic plasticity on ISR, we first review the phenomenon for the case of static synapses. Because we consider a postsynaptic neuron that receives balanced excitatory and inhibitory inputs, the resulting postsynaptic current has a mean of zero, independent of the presynaptic firing rate $f$. However, fluctuations about this mean grow monotonically with $f$. These fluctuations consititute a noisy input that is delivered to the postsynaptic neuron. Thus, we plot the mean firing rate as a function of $f$, over a biophysically reasonable range, in our analysis of the emergence of ISR. This is presented in Fig.~\ref{fig1}a.

{{The mean firing rate of the postsynaptic neuron, $\nu$, exhibits a pronounced minimum as a function of $f$: we will refer hereafter to the shape of this curve as the ISR ``well''.}} The fundamental mechanism underlying this non-monotonic behavior was described in detail in \cite{uzuntarla13}. We briefly summarize the main points here, and we visualize the spiking behavior of the postsynaptic neuron in Fig.~\ref{fig1}b-f at specific regions along the ISR curve marked in Fig.~\ref{fig1}a.

In region ``b'', the high average firing rate is due to what we call the ``initial condition effect''. This applies when the amplitude of the fluctuations in the noisy input into the postsynaptic neuron is very small. Depending on the randomly-chosen initial condition, the neuron is attracted to either the spiking state or the resting state. It then remains there, because the background fluctuation level is too low at such $f$ values to kick the system into the other state. This initial condition effect can be seen in Fig.~\ref{fig1}b, which shows that the neuron remains in its initial state over the course of each trial. Given the manner in which the initial condition is selected (described above), the neuron is most likely to be initiated within the basin of attraction of the spiking state \cite{TuckwellPhysRevE09,uzuntarla13}. Therefore, the mean firing rate value calculated over many trials is high in this region of the ISR curve \footnote{The value of $\nu$ can be predicted based on how the parameter space from which the initial condition is selected is proportioned with respect to the basins of the stable limit cycle and equilibrium. See \cite{uzuntarla13}.}.

In region ``c'', {{the postsynaptic current fluctuations induced by background activity are slightly larger}}. At these levels, the fluctutations become effective at changing the neuron's trajectory from the spiking to the resting state. However, they are not large enough to cause a change from the resting state to the spiking state.
This effect, which we call the “trapping effect”, is due to
%{\color {red}{This so-called “trapping effect” is due to}} 
the dynamical structure of the neuron model as described in \cite{uzuntarla13}. The resulting activity patterns of the postsynaptic neuron are shown in Fig.~\ref{fig1}c. In almost all trials, the neuron stops firing within the time interval shown. Thus, the contribution of the spiking state to $\nu$ decreases, resulting in the falling phase of the ISR curve. An extreme case of the trapping effect is seen in Fig.~\ref{fig1}d, in which the neuron stops firing almost immediately in every trial. Thus, $\nu$ effectively falls to zero; see region ``d'' in Fig.~\ref{fig1}a.

For values of $f$ in regions ``e'' and ``f'' of Fig.~\ref{fig1}a, the postsynaptic current fluctuations become strong enough to change the state of the neuron bidirectionally, i.e., from spiking to resting and vice versa. Thus the neuron exhibits on-off bursts of tonic firing as seen in Fig.~\ref{fig1}e. We call this the ``kickout effect'', and the ``kickouts'' (and ``kick-ins") occur more frequently as $f$ increases, resulting in the increase in the average firing rate $\nu$ as seen in Fig.~\ref{fig1}a; compare also Figs.~\ref{fig1}e and f. Finally, for $f>1000\, \mathrm{Hz}$, $\nu$ eventually saturates at a value somewhat higher than the spiking rate of an isolated, noise-free spiking neuron, which is approximately $58\, \mathrm{Hz}$ (not shown, but see \cite{uzuntarla13}).

\subsection{ISR with depressing synapses}

\begin{figure}
\includegraphics[scale=0.4]{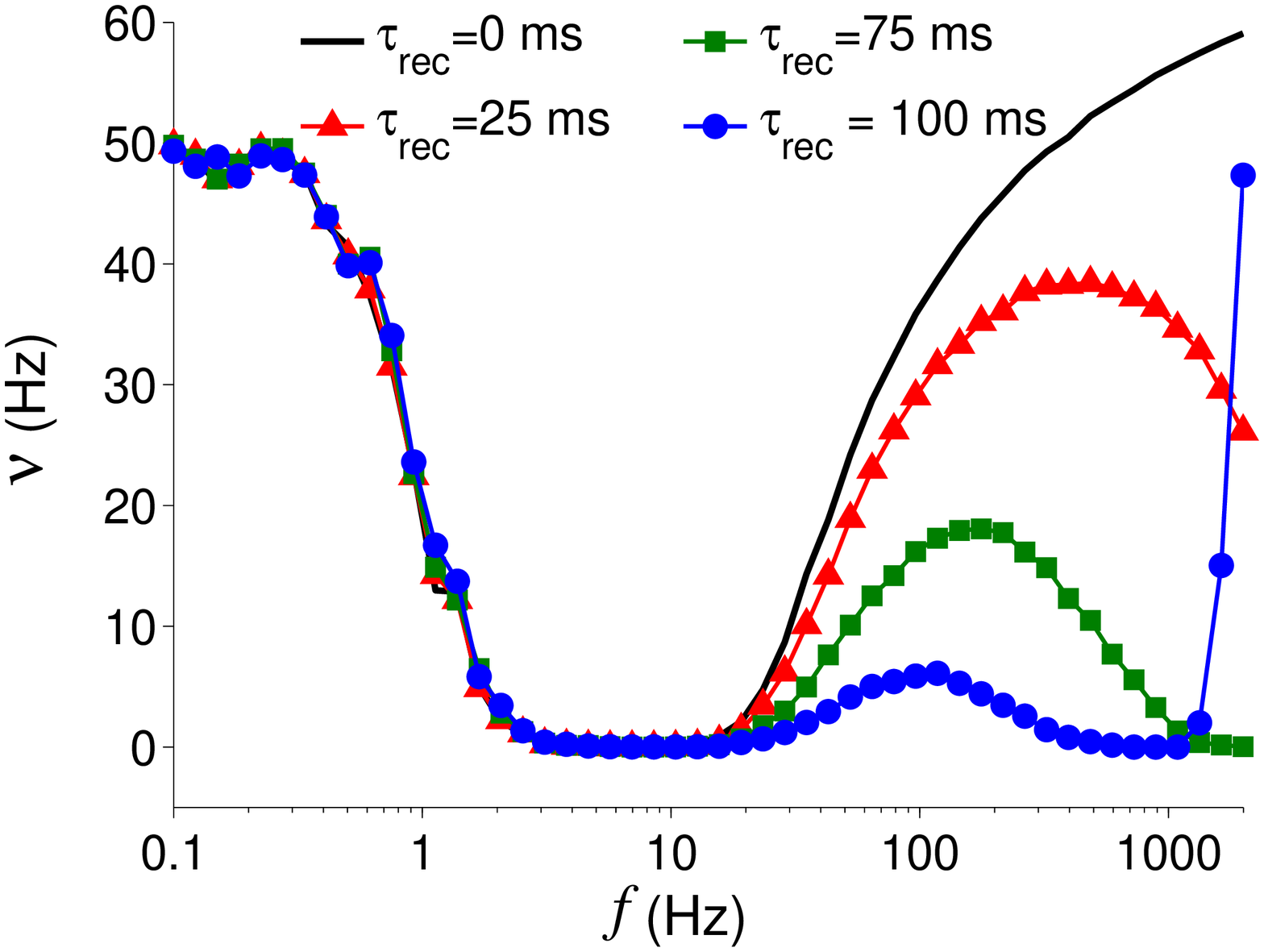} \\
(a) \\
\includegraphics[scale=0.4]{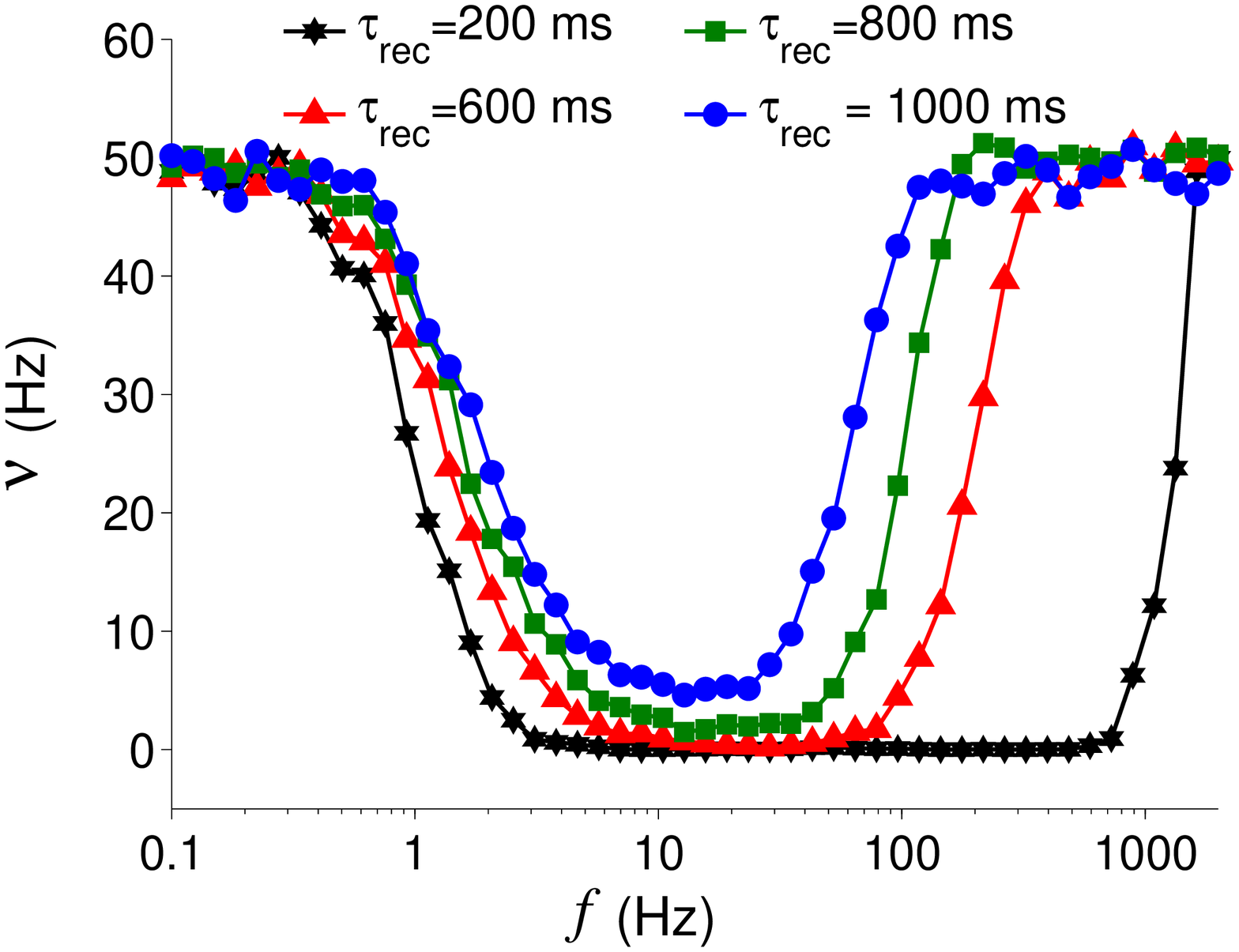} \\
(b)
\caption{\label{STDfigure}(Color online) Changes in the ISR behavior of a postsynaptic neuron that receives background activity through depressing synapses. The level of short-term depression at the synapses is controlled by the recovery time constant $\tau_{rec}$. Here, $\tau_{fac}=0\, \mathrm{ms}$, indicating the absence of synaptic facilitation. The figures show the average firing rate $\nu$ versus the presynaptic firing rate $f$ for relatively low \textbf{(a)} and high \textbf{(b)} levels of depression. Other synapse model parameters are as in Fig.~\ref{fig1}.}
\end{figure}

Next, we investigate the effects of short-term plasticity on the emergence of ISR. We begin by fixing the facilitation time constant $\tau_{fac}$ at zero in order to consider only the effects of STD. We then compute $\nu$ vs. $f$ for various values of the synaptic depression control parameter $\tau_{rec}$. The results are presented in Fig.~\ref{STDfigure},
which shows the effects of STD at low (panel (a)) and high levels (panel (b)). We see that STD mainly influences $\nu$ at high $f$ values. More precisely, for values of $\tau_{rec} \leq100\, \mathrm{ms}$, the right side of the ISR curve found in the case of static synapses ($\tau_{rec}=0\, \mathrm{ms}$; {{solid curve}}) tends to shift lower as $\tau_{rec}$ increases, with a more pronounced effect as $\tau_{rec}$ approaches $100\, \mathrm{ms}$. Nevertheless, for $\tau_{rec}$ around $100\, \mathrm{ms}$, $\nu$ suddenly increases for very large values of $f$, resulting in ISR behavior again. Panel (b) shows that for values of $\tau_{rec}>100\, \mathrm{ms}$, ISR is present but the width and depth of the ISR well is modulated by $\tau_{rec}$. In particular, the range of small $\nu$ values shrinks as $\tau_{rec}$ increases, and the lowest values of $\nu$ increase for very large values of $\tau_{rec}$. Moreover, for large $f$, $\nu$ saturates and attains a value similar to what is observed in the region of low $f$. Note that this is different from the static synapse case: compare the right sides of the ISR curves in panel (b) with the $\tau_{rec}=0\, \mathrm{ms}$ {{solid curve}} in panel (a).\\

\subsection{ISR in the presence of competing short-term depression and facilitation}

\begin{figure}
\includegraphics[scale=0.4]{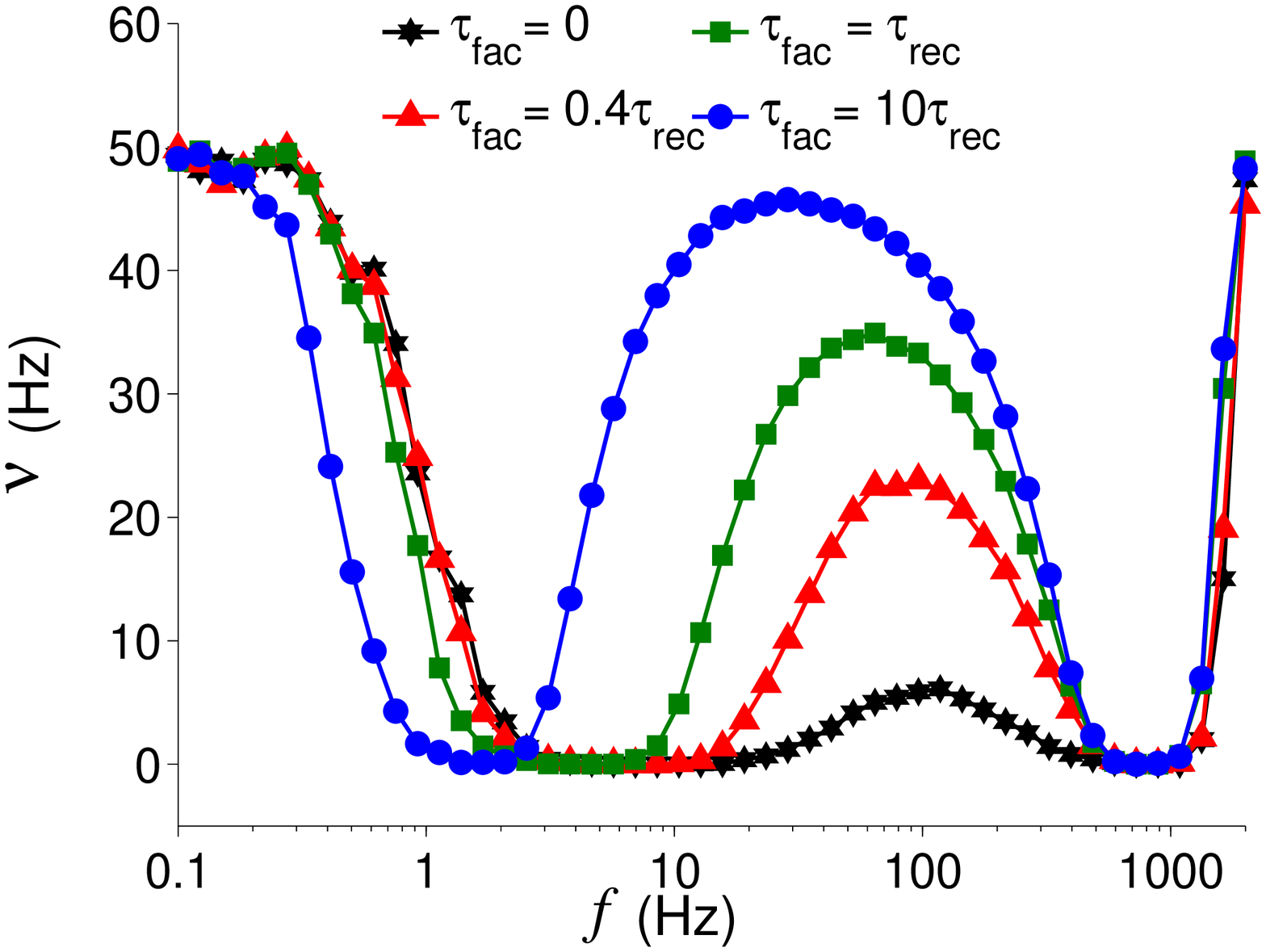} \\
(a) \\
\includegraphics[scale=0.4]{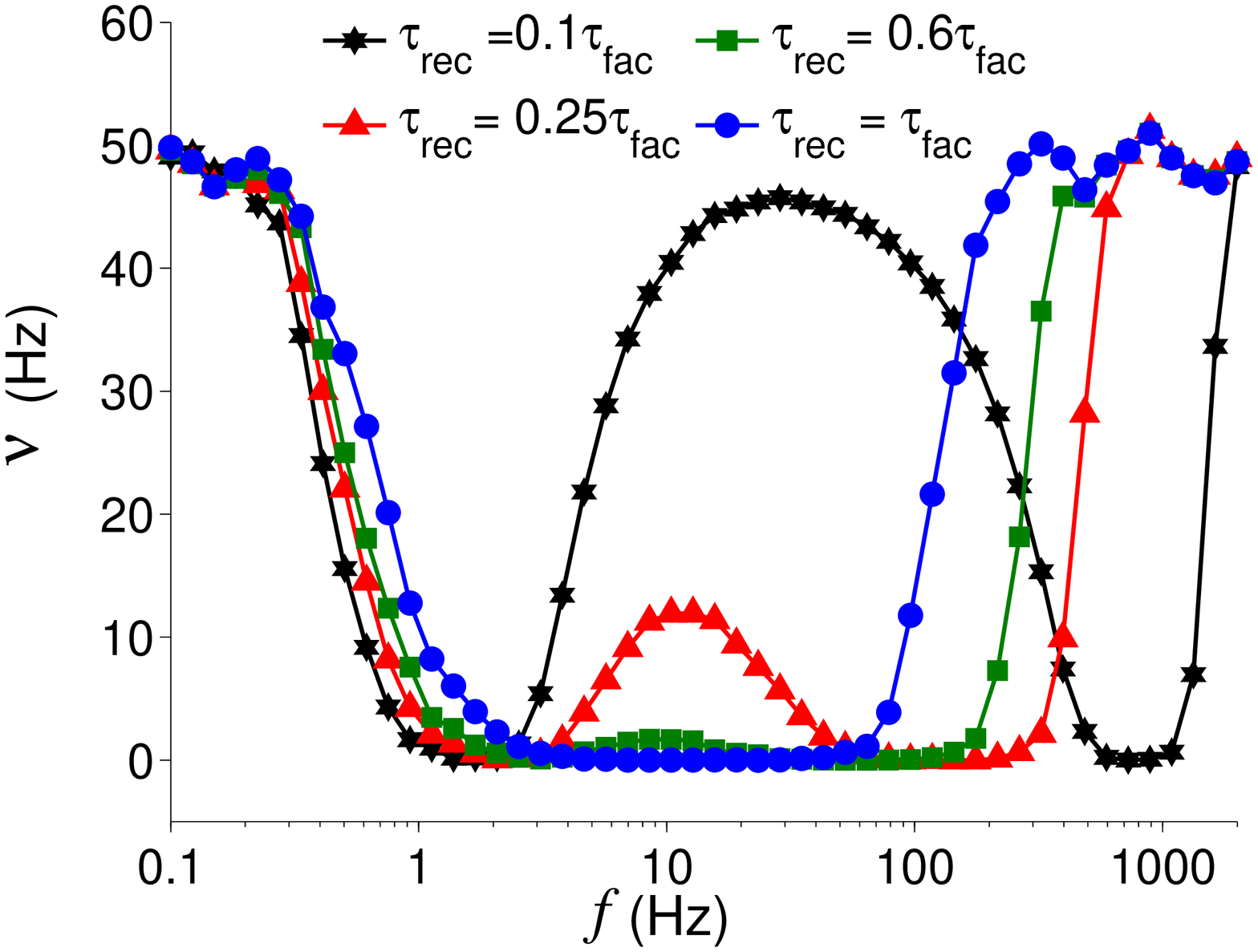} \\
(b)
\caption{\label{STDSTFcompetition}(Color online) The influence of both short-term depression and facilitation on ISR. \textbf{(a)} $\nu$ versus $f$ for various values of $\tau_{fac}$ for a fixed level of STD at synapses defined by $\tau_{rec}=100\, \mathrm{ms}$. \textbf{(b)} $\nu$ versus $f$ for various values of $\tau_{rec}$ for a fixed level of STF at synapses defined by $\tau_{fac}=1000\, \mathrm{ms}$ . Other synapse model parameters are as in Fig.~\ref{fig1}.}
\end{figure}

In actual synapses, both STD and STF can coexist, resulting in a nontrivial postsynaptic response \cite{Fitzpatrick2001,Ma2012,Flores2015,Cho13042011,zuckerARP02}.  Therefore, we may expect that the competition between these mechanisms {{could}} influence the ISR phenomenon significantly. To examine this effect, we first fixed the level of depression by setting the recovery time constant $\tau_{rec}$ to $100\, \mathrm{ms}$
%, a moderate level near the transition from the absence of ISR to its presence
(see Fig.~\ref{STDfigure}a). Then we systematically varied the facilitation via $\tau_{fac}$ to investigate the implications of the competition between STF and STD on ISR. The results are shown in Fig.~\ref{STDSTFcompetition}a.
We see that as $\tau_{fac}$ increases, the local maximum of $\nu$ located around $f=100\, \mathrm{ms}$ increases and shifts to lower values of $f$. This increase in $\nu$ at mid frequencies is more pronounced for large values of $\tau_{fac}$, and we see the emergence of two clearly separated wells, that is, double ISR (DISR). We see that STF affects the width and location of the low frequency well, making it narrower and shifting it towards lower $f$ as $\tau_{fac}$ increases. However, it does not affect the location of the high frequency well.

If we fix the level of facilitation instead at $\tau_{fac}=1000\, \mathrm{ms}$, where the DISR is most obvious in our previous analysis (with $\tau_{rec}=100\, \mathrm{ms}$), and varied the level of depression by tuning $\tau_{rec}$. The results are shown in Fig.~\ref{STDSTFcompetition}b. We see that the height of the local maximum at mid $f$ values decreases as $\tau_{rec}$ increases, and at the same time the high frequency wells in the DISR curves shift to lower $f$ values. For $\tau_{rec} > 600\, \mathrm{ms}$, the central peak disappears and the DISR wells merge into a wide single ISR well. Note that the low frequency wells of the DISR curves are not significantly affected by the level of depression. Thus, our findings for the case of competing synapses demonstrate that STF favors the appearance of DISR, whereas STD tends to work oppositely, favoring single ISR behavior.\\

\subsection{Mechanism of ISR and DISR with Dynamic Synapses}

\begin{figure}
\hspace*{-0.5cm}\includegraphics[scale=0.40]{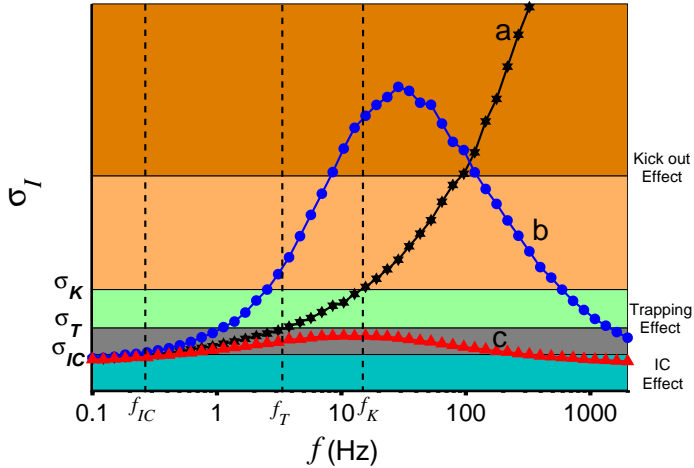}
\caption{\label{fluctfig} (Color online) A qualitative depiction of the standard deviation $\sigma_I$ of the synaptic current versus the presynaptic firing rate $f$ for static (curve {\em a}), competing (curve {\em b}) and depressing (curve {\em c}) synapses. The vertical axis is schematic, and the three horizontal lines marked $\sigma_{IC}$, $\sigma_{T}$, and $\sigma_{K}$ separate regions (colored as in Fig.~\ref{fig1}a) that correspond to the different dynamic mechanisms described in the main text. The vertical dashed lines are the same as those in Fig.~\ref{fig1}a. Synaptic parameter sets for curve ({\em a}) $\tau_{rec}=0\, \mathrm{ms}$, $\tau_{fac}=0\, \mathrm{ms}$, curve ({\em b}) $\tau_{rec}=100\, \mathrm{ms}$, $\tau_{fac}=1000\, \mathrm{ms}$, curve ({\em c}) $\tau_{rec}=1000\, \mathrm{ms}$, $\tau_{fac}=0\, \mathrm{ms}$. Other synapse model parameters are as in Fig.~\ref{fig1}.}
\end{figure}

A qualitative understanding of the results reported above can be gained by considering the nature of the postsynaptic current fluctuations and their dependence on the presynaptic firing rate $f$. In the case of static synapses, the standard deviation of the postsynaptic current $\sigma_I$ rises monotonically with $f$, since the presynaptic firing events are Poisson-distributed. This is shown in Fig.~\ref{fluctfig}, curve {\em a}.

In contrast, when the synapses feature short-term plasticity, the postsynaptic current fluctuations exhibit non-monotonic behavior. This is due to synaptic depression, which is strongest for higher presynaptic firing rates. Under such circumstances, neurotransmitter resources are quickly depleted, and the postsynaptic current is attenuated and eventually eliminated. Correspondingly, the current fluctuations also decrease to zero, as can be seen in Fig.~\ref{fluctfig}, curves  {\em b} and  {\em c}. Including facilitation leads to a similar unimodal curve, but with the central region shifted higher (curve  {\em b}). This is because facilitation is strongest for intermediate presynaptic firing rates, and leads to more reliable synaptic transmission. Thus the postsynaptic current is enhanced, and it reflects more closely the fluctuations in the Poissonian presynaptic firing pattern, leading to an increase in $\sigma_I$. However, this effect can lead to particularly efficient transmitter depletion for high presynaptic firing rates, and hence depression. Therefore, the postsynaptic current and its fluctuations again decrease to zero for high $f$.

The horizontal lines labeled $\sigma_{IC}$, $\sigma_{T}$, and $\sigma_{K}$ in Fig.~\ref{fluctfig} schematically indicate the ranges of $\sigma_I$ that correspond to the different dynamic mechanisms listed above. To determine these noise levels, we follow the monotonically-increasing curve {\em a} in Fig.~\ref{fluctfig} for the case of static synapses as $f$ increases, and compare to the ISR curve in Fig.~\ref{fig1}a. By using the intersections of the vertical lines corresponding to $f_{IC}$, $f_{T}$, and $f_{K}$ (that separate different postsynaptic neuron behaviors as defined in Fig.~\ref{fig1}a) with $\sigma_{I}$ curve {\em a} for static synapses, we respectively determine $\sigma_{IC}$, $\sigma_{T}$, and $\sigma_{K}$. Values of $\sigma_I$ below the horizontal line at $\sigma_{IC}$ correspond to the initial condition effect, and to the region marked ``b'' in Fig.~\ref{fig1}a. For $\sigma_I$ between the horizontal lines at $\sigma_{IC}$ and $\sigma_{T}$, the trapping effect is observed. As $\sigma_I$ increases through the region between these lines, trapping becomes increasingly probable, and there is a transition from the initial condition effect to full trapping. This corresponds to the decrease in $\nu$ observed in region ``c'' of Fig.~\ref{fig1}a. For $\sigma_I$ between the horizontal lines 
at $\sigma_{T}$ and $\sigma_{K}$, trapping occurs in essentially all trials, and $\nu$ decreases to zero; see region ``d'' of Fig.~\ref{fig1}a. Finally, current fluctuations above the horizontal line at $\sigma_{K}$ are large enough to cause the kickout effect, and lead to the increasing phase of the ISR curve seen in regions ``e'' and ``f'' of Fig.~\ref{fig1}a.

This schematic description provides a heuristic understanding of how DISR arises. If the unimodal variation of $\sigma_I$ crosses the schematic horizontal lines shown in Fig.~\ref{fluctfig}, then we can expect a plot of $\nu$ versus $f$ to display the following features, in order as $f$ increases, as can be observed (for example) in the $\tau_{fac}=1000\, \mathrm{ms}$ data of Fig.~\ref{STDSTFcompetition}(a): First, $\nu$ will be high, corresponding to the initial condition effect and values of $\sigma_I$ below the the $\sigma_{IC}$ horizontal line. Then $\nu$ will decrease to zero, corresponding to the trapping effect and values of $\sigma_I$ between the horizontal lines at $\sigma_{IC}$ and $\sigma_{K}$. After this, $\nu$ will increase due to the kickout effect as $\sigma_I$ crosses the upper horizontal line at $\sigma_{K}$. As $f$ continues to increase, kickout effects become more frequent, $\nu$ increases, and $\sigma_I$ attains its maximum value. After this, $\sigma_I$ decreases with increasing $f$, kickout events become less frequent, and $\nu$ decreases. As $\sigma_I$ returns to the region between the horizontal lines, trapping events once again occur, and $\nu$ again decreases to zero. Finally, for the highest values of $f$, $\sigma_I$ returns to low values, and the re-emerging initial condition effect causes $\nu$ to rise once again. Therefore, the plot of $\nu$ versus $f$ displays two wells, and hence DISR.

Depending on parameters, especially for cases with little or no facilitation, it is possible for the unimodal $\sigma_I$ curve to cross only the lowest horizontal line, as in curve {\em c} in Fig.~\ref{fluctfig}. In this case, a similar scenario arises, but without the occurrence of kickout effects. Thus a plot of $\nu$ versus $f$ would begin high (initial condition effect), decrease, perhaps to near zero (trapping effect), and then increase again (initial condition effect), yielding a single-well ISR curve. This behavior is seen in Fig.~\ref{STDfigure}b and in Fig.~\ref{STDSTFcompetition}b for $\tau_{rec}>600\, ms$. Note also that the symmetry in this heuristic perspective explains the similarity of the values of $\nu$ observed at low and high values of $f$ in these figures, as both are due to the same initial condition effect.\\

\begin{figure*}[pbth]
\begin{centering}
\hspace*{-0.5cm}\includegraphics[scale=0.40]{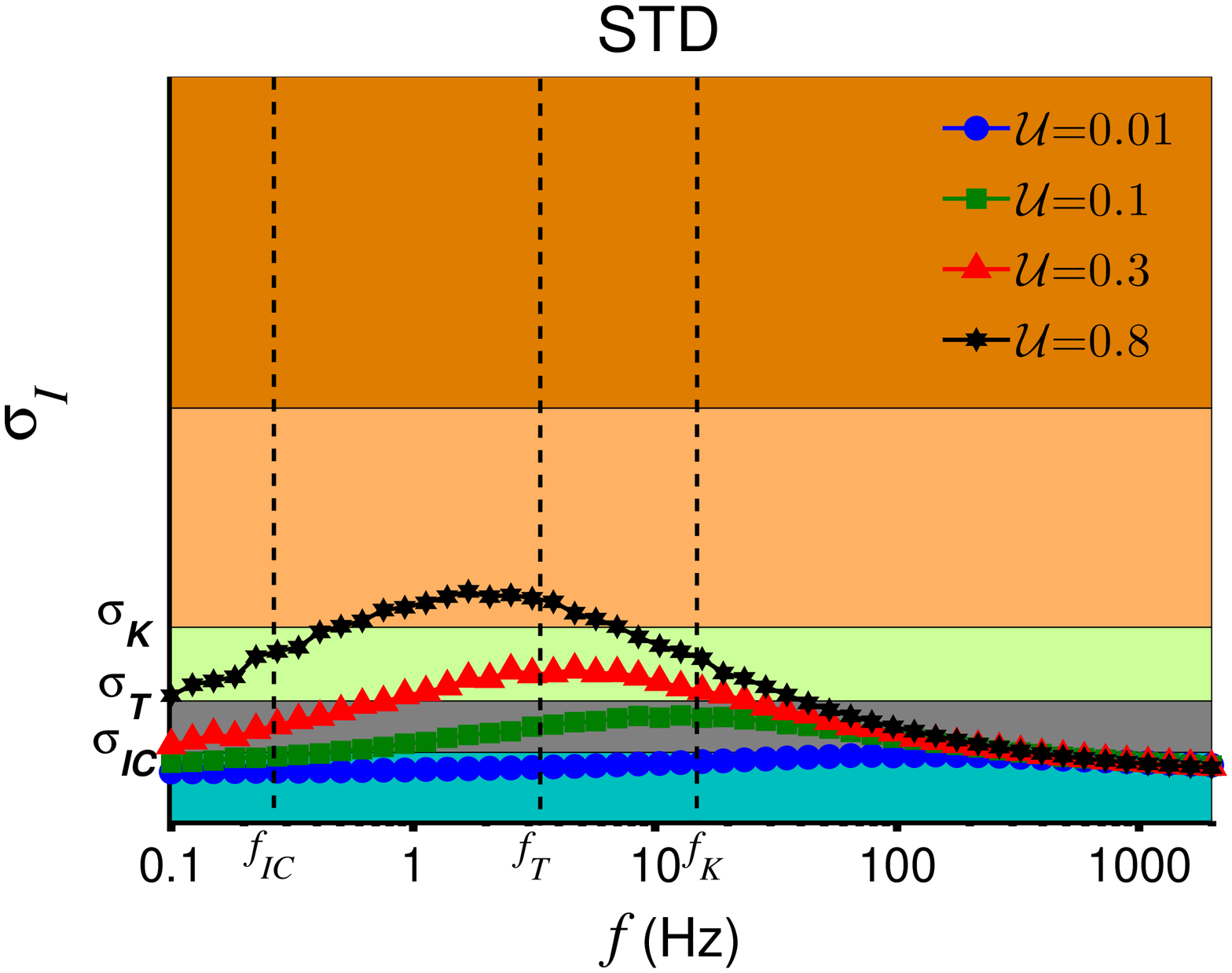}\includegraphics[scale=0.40]{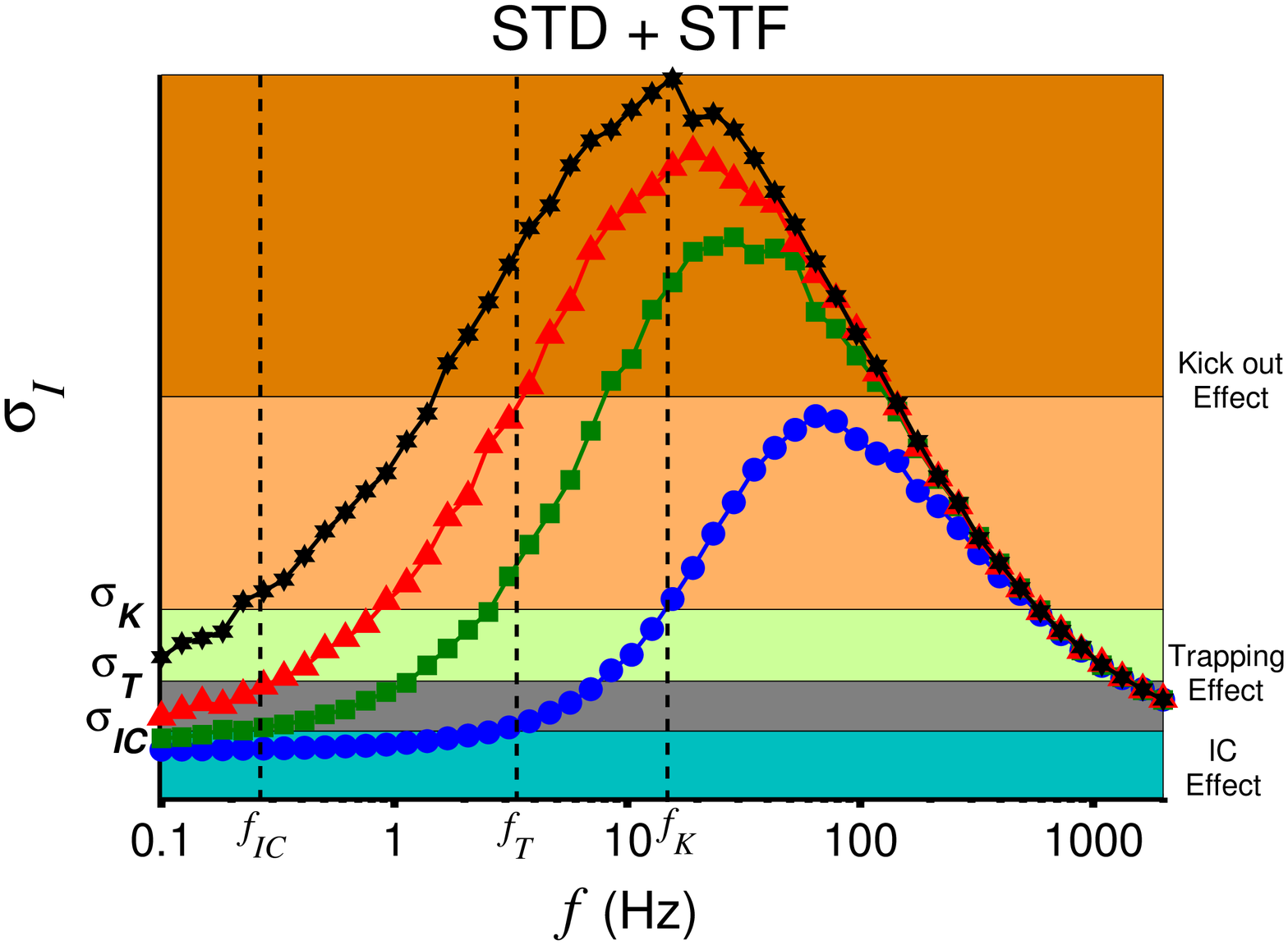}
\par\end{centering}
\begin{centering}
(a)~~~~~~~~~~~~~~~~~~~~~~~~~~~~~~~~~~~~~~~~~~~~~~~~~~~~~~~~~~~~~~~~~~~~~~~~~~~~~~~~
(b)~~~~
\par\end{centering}
\begin{centering}
\hspace*{-0.5cm}\includegraphics[scale=0.40]{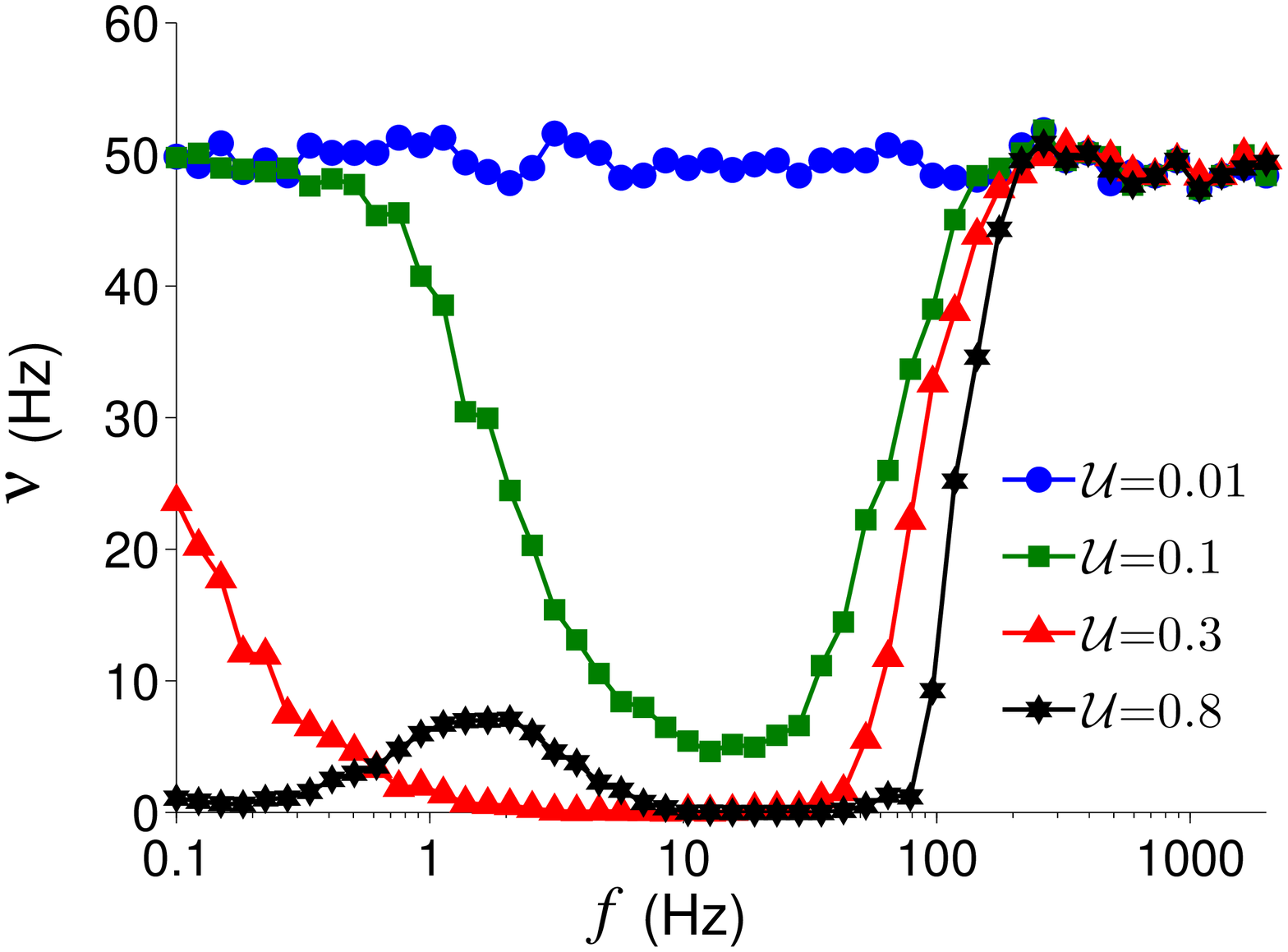}\includegraphics[scale=0.40]{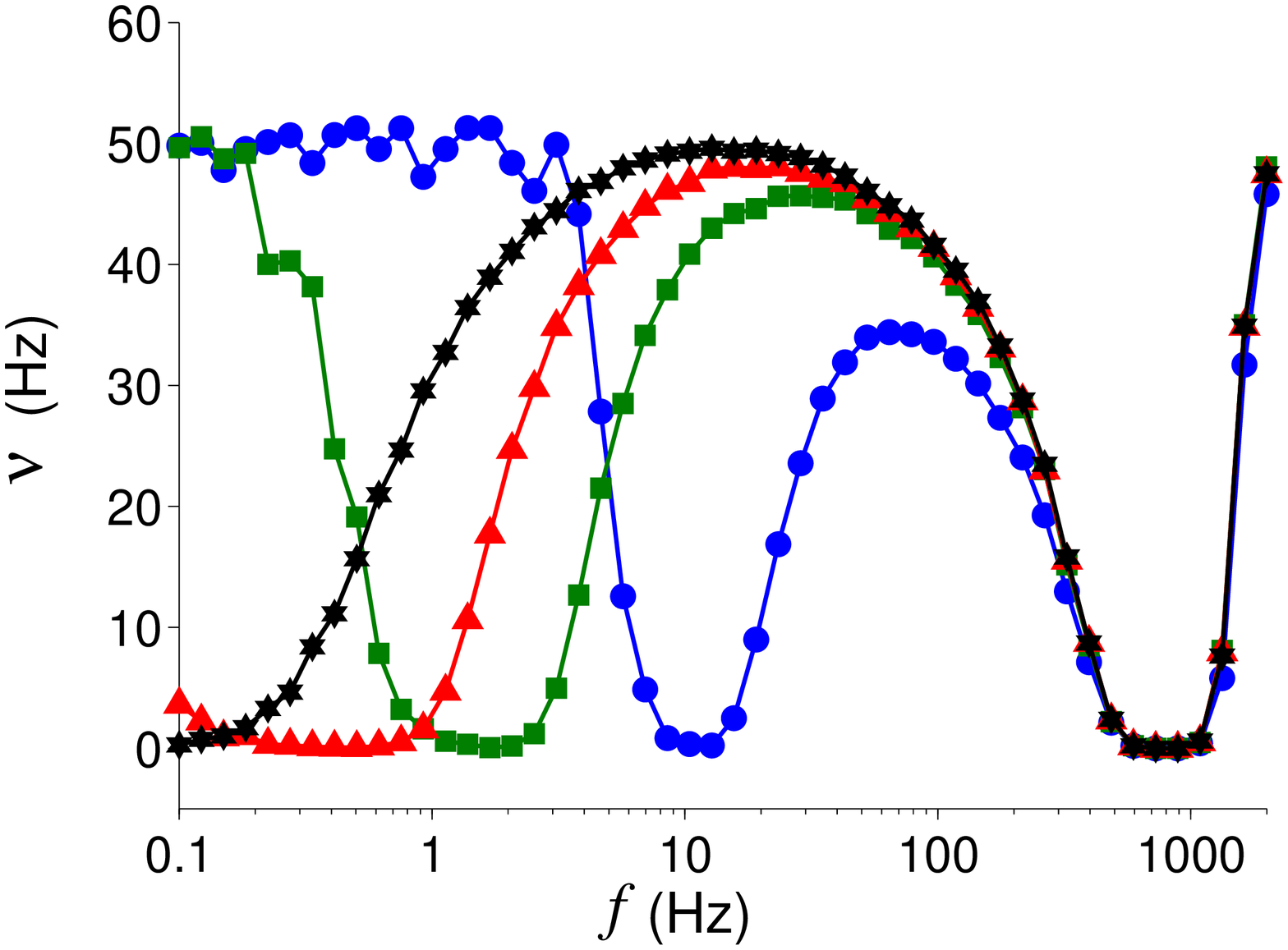}
\par\end{centering}
\begin{centering}
(c)~~~~~~~~~~~~~~~~~~~~~~~~~~~~~~~~~~~~~~~~~~~~~~~~~~~~~~~~~~~~~~~~~~~~~~~~~~~~~~~~
(d)~~~~
\par\end{centering}
\caption{(Color online) Effects of the baseline fraction of released neurotransmitter $\mathcal{U}$ on synaptic current fluctuations and the postsynaptic mean firing rate. \textbf{(a)} Variation of $\sigma_{I}$ as a function of $f$ for different values of $\mathcal{U}$ when only STD is present, with $\tau_{rec}=1000\, \mathrm{ms}$ and $\tau_{fac}=0\, \mathrm{ms}$. \textbf{(b)} Similar analysis as in (a) for the case of competing STD and STF, with $\tau_{rec}=100\, \mathrm{ms}$ and $\tau_{fac}=1000\, \mathrm{ms}$. Note that the three schematic horizontal lines marked $\sigma_{IC}$, $\sigma_{T}$, and $\sigma_{K}$ separate regions {{(as in Fig.~\ref{fig1}a)}} that correspond to the different dynamic mechanisms described in the main text. {\textbf{(c)} Variation of $\nu$ as a function of $f$ for different values of $\mathcal{U}$ in the case of depressing synapses. This shows the transition from no ISR to a single-well ISR and then to a weak DISR as $\mathcal{U}$ increases. \textbf{(d)} Variation of $\nu$ as a function of $f$ for different values of $\mathcal{U}$. In this case, the plots show the transition from DISR to a single-well ISR when STD and STF are both present at synapses. \label{Uvar}}}
\end{figure*}

\subsection{The influence of $\mathcal{U}$ on ISR and DISR}

We now consider the effects of varying $\mathcal{U}$, which appears in Eq.~\ref{dynueq} and represents the fraction of available neurotransmitter resources that transitions to the active state upon the arrival of a presynaptic spike, when the synapse is in its resting state.
%It has been shown that $\mathcal{U}$ can critically control the dynamic behavior of synapses \cite{tsodyks2005course, Markram1998489}, a fact that has intriguing computational implications \cite{torresNC2007,mejiasCD08,mejias09,TorresKappen2013, Mongillo2012, mejias2011emergence}.
We follow the schematic description developed above and plot $\sigma_I$ versus $f$ for various values of $\mathcal{U}$, for the cases that we previously considered: synapses with just depression (Fig.~\ref{Uvar}a), and with both depression and facilitation (Fig.~\ref{Uvar}b).

We observe that variation of $\mathcal{U}$ results in similar effects on $\sigma_I$ in both cases, namely, an amplification of $\sigma_I$ at low and moderate $f$ values, while $\sigma_I$ does not change at high values of $f$. The reason for this latter effect is that at high frequencies, $\mathcal{U}$ becomes irrelevant because neurotransmitter resources rapidly deplete, especially if the recovery time is relatively long ($\tau_{rec}=1000\, \mathrm{ms}$ in Fig.~\ref{Uvar}a and $\tau_{rec}=100\, \mathrm{ms}$ in Fig.~\ref{Uvar}b). In contrast, at low and moderate $f$ values, there is more time for neurotransmitter resources to recover, and $\mathcal{U}$ has more significant effects. Thus we see that as $\mathcal{U}$ increases, the curves for the various cases change position in relation to the horizontal lines that indicate the onset of the initial condition, trapping, and kickout effects described above.

\begin{figure*}[pbth!]
\begin{centering}
\hspace*{-0.5cm}\includegraphics[height=5.8cm, width=9cm]{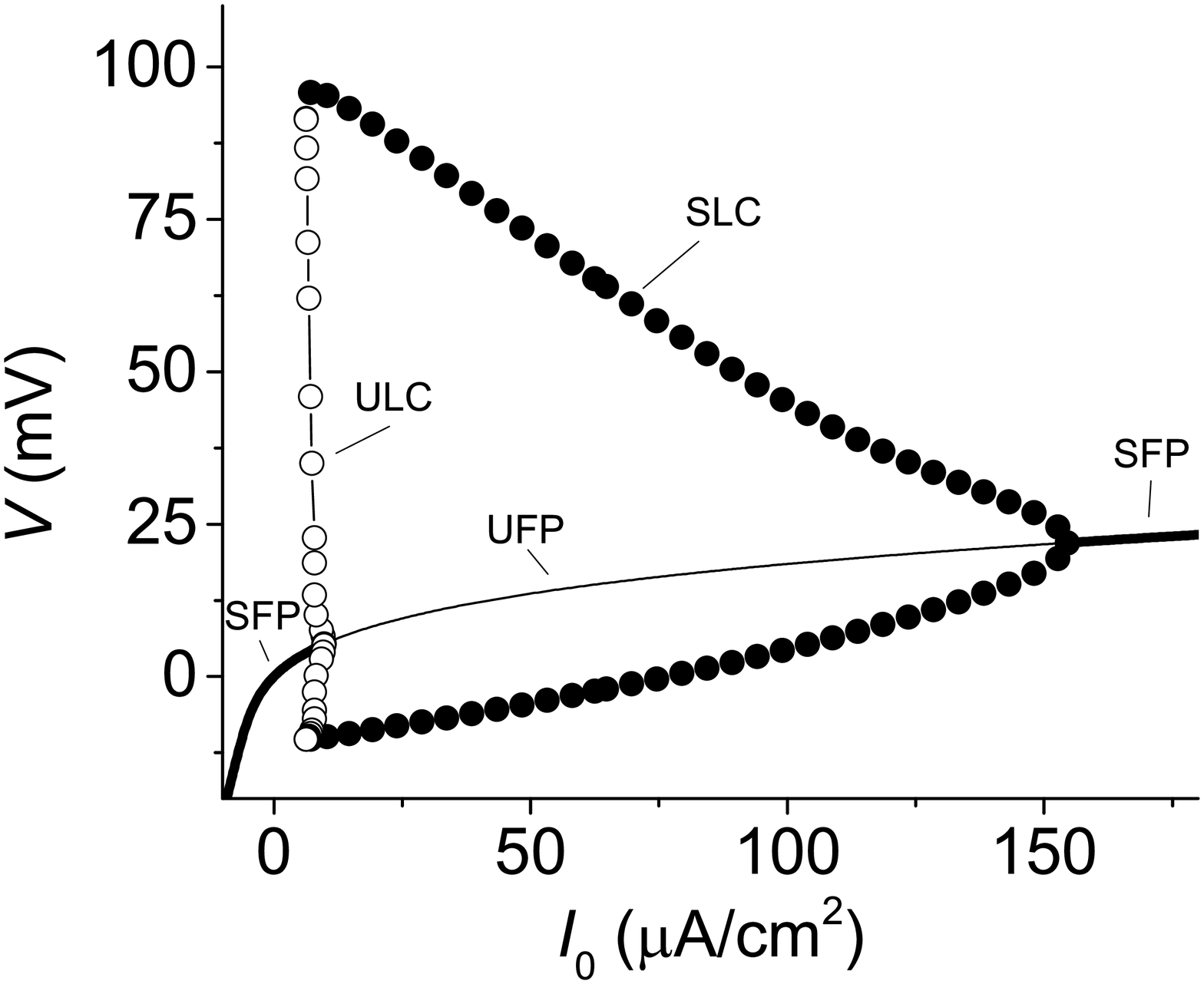}~\hspace*{-0.8cm}\includegraphics[height=5.8cm, width=9cm]{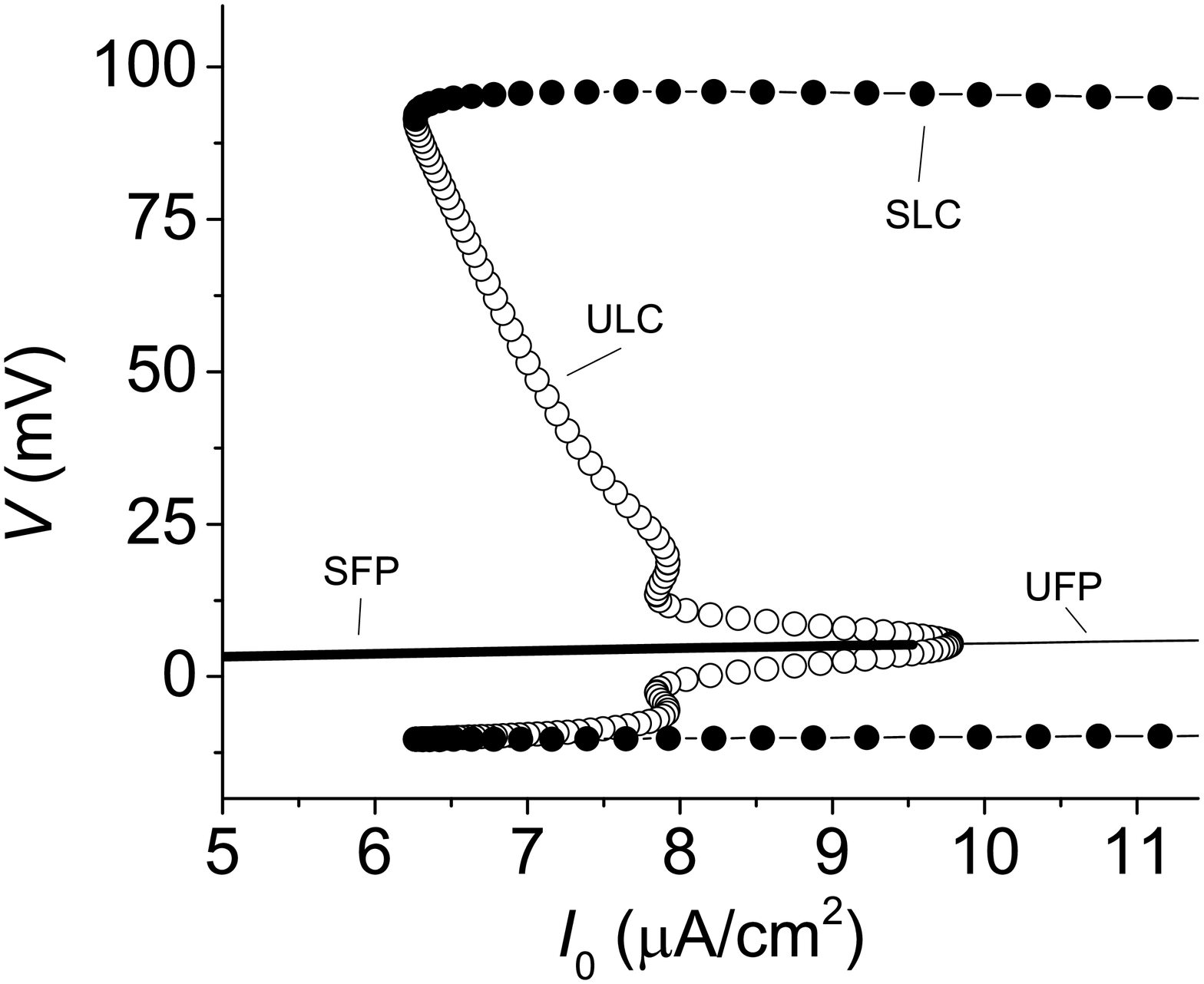}
(a)~~~~~~~~~~~~~~~~~~~~~~~~~~~~~~~~~~~~~~~~~~~~~~~~~~~~~~~~~~~~~~~~~~~~~~~~
(b)~~~~
\par\end{centering}
\caption{\label{bifs} Panel \textbf{(a)} shows the bifurcation diagram of the Hodgkin-Huxley neuron as a function of the bias current $I_0$. Thick (thin) solid lines represent stable (unstable) fixed points, marked SFP and UFP. Solid (open) circles represent the minimum and maximum values of the voltage during stable (unstable) spiking behavior. These are marked SLC and ULC. Panel \textbf{(b)} shows a magnification of panel (a), revealing the multistable region.}
\end{figure*}

More specifically, for the STD case, we see that increasing $\mathcal{U}$ leads to a transition, as shown in Fig.~\ref{Uvar}c, from no ISR {{(circle)}} to a single-well ISR {{(square and triangle)}} and then to a weak double-well ISR {{(star)}}. On the other hand, for the case with both STD and STF, we observe that increasing $\mathcal{U}$ results in the disappearance of the low frequency DISR well, thus leading to a transition from double ISR {{(circle and square)}} to a single-well ISR {{(star and triangle)}} as shown in Fig.~\ref{Uvar}d. Such transitions arise in both cases from the sensitivity of $\sigma_I$ to $\mathcal{U}$ at low frequencies. As seen in Fig.~\ref{Uvar}a and b, the left ends of the $\sigma_I$ vs. $f$ curves increase with increasing $\mathcal{U}$ at the low frequency end starting at $f=0.1\, \mathrm{Hz}$, while the high frequency ends of $\sigma_I$ vs. $f$ curves are not affected. For instance, in the case of only STD, the leftmost values for $\mathcal{U}=0.8$ lie close to the $\sigma_{T}$  horizontal line, indicating essentially full trapping. Accordingly, the corresponding $\nu$ values at $f=0.1\, \mathrm{Hz}$ decrease to zero as $\mathcal{U}$ increases, and the left ISR well gradually disappears. This observation also explains the transition from DISR to single-well ISR as $\mathcal{U}$ increases in the case of competing STD and STF (compare Fig.~\ref{Uvar}c and d).\\

\subsection{Role of neuronal excitability on ISR and DISR}

Finally, we examine the effect of changing $I_0$. ISR emerges when the system dynamics exhibits bistability that includes a stable fixed point (SFP), an unstable limit cycle (ULC), and a stable limit cycle (SLC) \cite{TuckwellPhysRevE09,uzuntarla13,Uzuntarlasolo13,Guo11,schmerl13}. For the Hodgkin-Huxley neuron, this situtation is present for a modest range of constant input current $I_0$. Fig.~\ref{bifs} illustrates the corresponding bifurcation diagram. Specifically, the bistable range extends  -- in the absence of other current inputs -- from approximately $I_0=6.26\,\mathrm{\mu A/cm^{2}}$, where the SLC and ULC are born by saddle-node bifurcation, to approximately $9.78\, \mathrm{\mu A/cm^{2}}$, where the ULC and SFP merge in a Hopf bifurcation (see panel (b)). At this point, the SFP loses stability becomes an unstable fixed point (UFP).

\begin{figure*}[ht!]
\begin{centering}
\hspace*{-0.5cm}\includegraphics[scale=0.4]{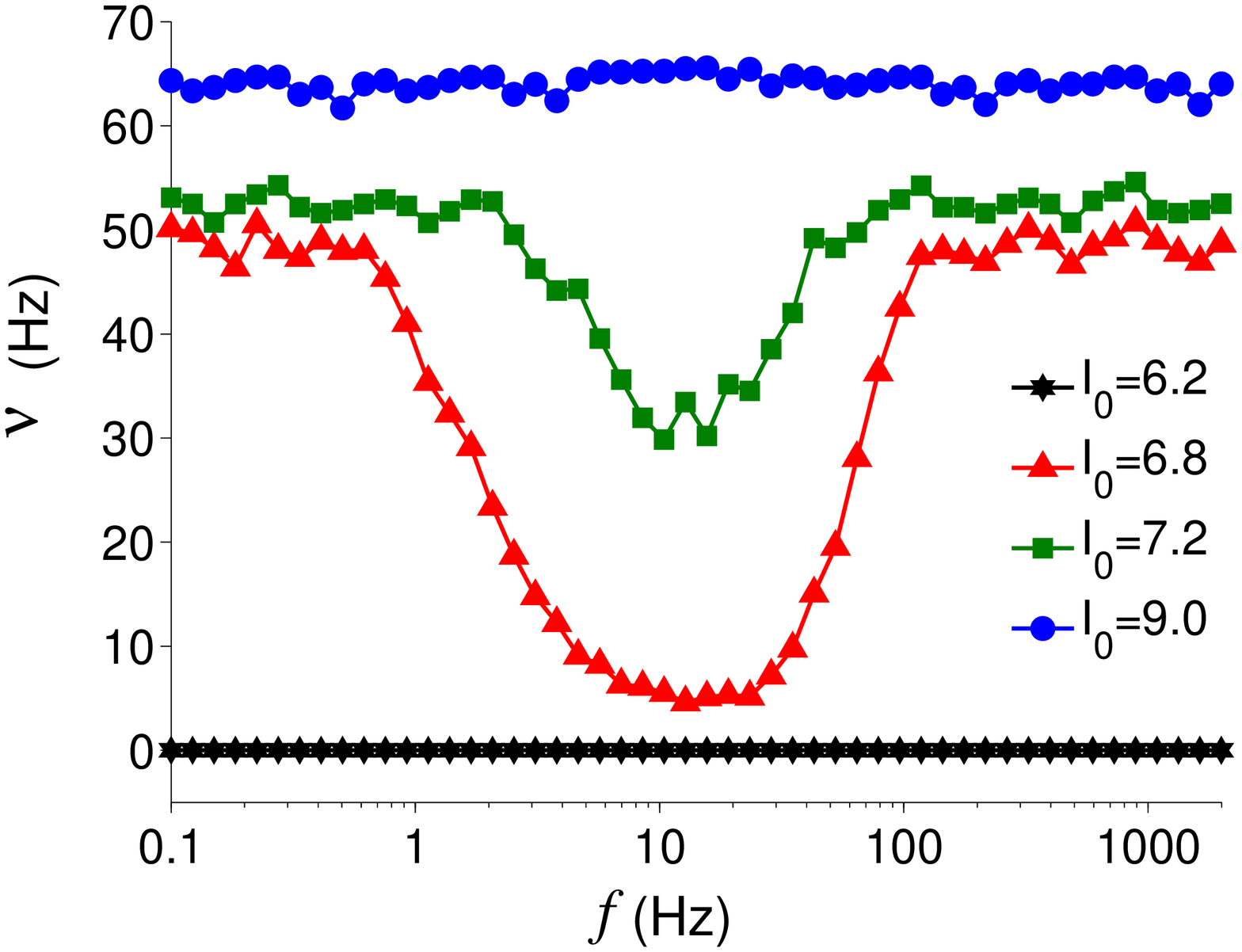}~\hspace*{-0.8cm}\includegraphics[scale=0.4]{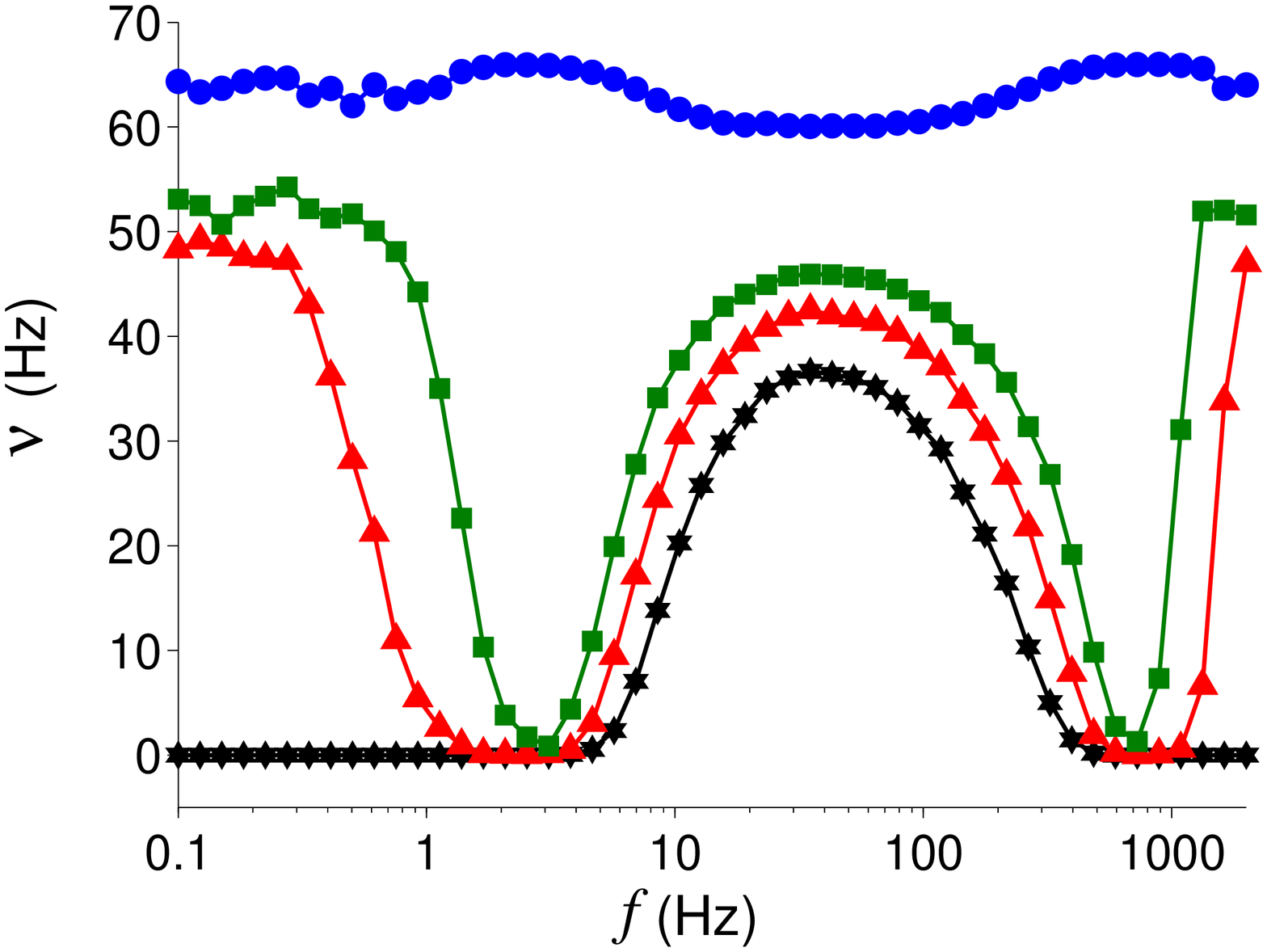}
(a)~~~~~~~~~~~~~~~~~~~~~~~~~~~~~~~~~~~~~~~~~~~~~~~~~~~~~~~~~~~~~~~~~~~~~~~~~~~
(b)~~~~
\par\end{centering}
\caption{\label{varycurrentfigure}(Color online) The effect of the excitability control parameter $I_{0}$ on the behavior of ISR and DISR. \textbf{(a)} $\nu$ vs. $f$ for different values of $I_{0}$ in the case of purely depressing synapses ($\tau_{rec}=1000\, \mathrm{ms}$ and $\tau_{fac}=0\, \mathrm{ms}$) that induce ISR at $I_{0}\sim6.8\,\mathrm{\mu A/cm^{2}}$. \textbf{(b)} Similar analysis as in (a) for the case of competing STD and STF mechanisms ($\tau_{rec}=100\, \mathrm{ms}$ and $\tau_{fac}=400\, \mathrm{ms}$) that result in DISR for $I_{0}\sim6.8\, \mathrm{\mu A/cm^{2}}$. Other synapse parameters are as in Fig.~\ref{fig1}.}
\end{figure*}

The ISR and DISR results reported above were obtained with $I_{0}=6.8\,\mathrm{\mu A/cm^{2}}$. In the neuronal context, the SFP and SLC correspond to resting and spiking behavior, respectively, and the ULC mediates the boundary between the basins of attraction of these two stable states \footnote{More precisely, the boundary is the closure of the stable manifold of the ULC.}. Within the bistable regime, the SFP and SLC are not significantly affected by variation in the excitability control parameter $I_{0}$, but it does significantly change the ULC, and correspondingly, the basin boundary \cite{Rowat2007,uzuntarla13}. In particular, the basin of the SFP dramatically decreases in size as the ULC converges to the SFP as the Hopf bifurcation is approached.
%Accordingly, the required levels of background fluctuations for the occurance of above mentinoned dynamic mechanism decrease because   
Therefore, we investigated the emergence of ISR and DISR for various values of $I_{0}$ near and within the bistable range. The results of our analysis are shown in Fig.~\ref{varycurrentfigure} for the case of (a) ISR with STD and (b) DISR with both STD and STF.

We see that increasing $I_{0}$ results, in both cases, in an overall increase {{of}} $\nu$ over the whole range of $f$. This is as expected, since as the basin of the SFP shrinks to zero for increasing $I_0$, a neuron in the resting state is more and more likely to be kicked into the basin of the spiking state (i.e., the SLC) and remain there \cite{Rowat2007,uzuntarla13}. This increase of $\nu$ is most dramatic where the minima of the ISR or DISR curves are located, leading to a gradual flattening of the curve and the suppression of the ISR or DISR effect. It is interesting to note that for $I_0=6.2\, \mathrm{\mu A/cm^{2}}$, only the SFP exists, and the system is not bistable (in the absence of other current inputs). Accordingly, Fig.~\ref{varycurrentfigure}(a) shows that for this case, $\nu=0\, \mathrm{Hz}$ over the entire range of $f$. However, in panel (b), with both STD and STF, we see that $\nu$ is not zero for intermediate values of $f$ for this case. The background activity triggers STF in this range, increasing $I_{syn}$ such that the overall current input effectively places the neuron into the bistable regime where spiking behavior can occur. The firing frequency $\nu$ decreases back to zero for even higher frequencies due to the increasing STD, which essentially cancels this effect.

\section{DISCUSSION}

The aim of this work was to examine a neuron subject to a large number of presynaptic inputs, where presynaptic events occur stochastically. It has been shown that this source of noisy input gives rise to ISR, and here we focused on the effects of dynamic synapses that exhibit short-term facilitation and depression. The key observation is that these synaptic dynamics lead to a non-monotonic relationship between the fluctuations in the postsynaptic current and the rate at which presynaptic events occur. Then, with the qualitative understanding of the neuronal dynamics for varying values of noise, we developed a heuristic understanding of how ISR rises, finding that, with the proper biophysical conditions, double ISR could also be found.

Recent experimental works indeed reported the existence of ISR in individual neurons \cite{paydarfar_noisy_2006, Buchin2016}, but the functional implications of such phenomena remain unknown partly because the experimental conditions for the observation of ISR have not been systematically explored. In this regard, our study argues in favor of the emergence of ISR by short-term synaptic plasticity, and suggests that experimental manipulation of the mechanisms underpinning such plasticity could lead to observable changes in biophysical correlates for ISR in real neurons \cite{Purves, Nagy}.

Although it is widely assumed that information is processed by spikes, the silent periods characterizing ISR may also have functional relevance. For instance, ISR may play a role in shortening the periods of anomalous working memory \cite{Dipoppa2013}. It has also been suggested that Purkinje cells involved
in cerebellar computation could use ISR to switch among different operating
regimes depending on input current fluctuations \cite{Buchin2016}. This leads to the speculation that ISR
mechanisms may generally provide a way for a neuron (or a population thereof) to be modulated by noisy inputs either without, or perhaps in conjunction with, stimulus-evoked signals. In this fashion, the low average firing rate of postsynaptic neurons in the ISR well could allow filtering of irrelevant information. This would facilitate neuronal tuning insofar as neurons could selectively process relevant information
arriving from different inputs. And double ISR may provide a mechanism in this direction that extends the possible biophysical conditions for which such selective tuning and multiplexing can be achieved.

In our study, we used the Tsodyks-Markham model of dynamic synapses, but we expect that any model that captures biophysical synaptic depression would lead to similar results. On the other hand, other biophysical factors can regulate synaptic activity in complex ways, such as in two-way synapse-astrocyte interactions \cite{PittaBrunel2016, DePitta2016}. Thus it might be interesting to explore the role of the mechanisms involved in such interactions (i.e., calcium dynamics) on the emergence of ISR and DISR. It may also be possible to extend our study to investigate ISR in simpler neuron models, such as two-dimensional models, to allow for theoretical analysis to determine, for example, the dependence of critical noise levels on different neuronal behaviors.

\section{Acknowledgments}

MU acknowledges financial support from the Scientific and Technological Research Council of Turkey (T\"{U}B\.{I}TAK) B\.{I}DEB-2219 Postdoctoral Research program, and the hospitality of the Krasnow Institute for Advanced Study at George Mason University. We thank Theodore Dumas and Ali \c{C}al{\i}m for helpful discussions. We also thank an anonymous reviewer for extensive and very helpful comments.

% Create the reference section using BibTeX:
%\bibliography{bibtotal}

\begin{thebibliography}{63}%
\makeatletter
\providecommand \@ifxundefined [1]{%
 \@ifx{#1\undefined}
}%
\providecommand \@ifnum [1]{%
 \ifnum #1\expandafter \@firstoftwo
 \else \expandafter \@secondoftwo
 \fi
}%
\providecommand \@ifx [1]{%
 \ifx #1\expandafter \@firstoftwo
 \else \expandafter \@secondoftwo
 \fi
}%
\providecommand \natexlab [1]{#1}%
\providecommand \enquote  [1]{``#1''}%
\providecommand \bibnamefont  [1]{#1}%
\providecommand \bibfnamefont [1]{#1}%
\providecommand \citenamefont [1]{#1}%
\providecommand \href@noop [0]{\@secondoftwo}%
\providecommand \href [0]{\begingroup \@sanitize@url \@href}%
\providecommand \@href[1]{\@@startlink{#1}\@@href}%
\providecommand \@@href[1]{\endgroup#1\@@endlink}%
\providecommand \@sanitize@url [0]{\catcode `\\12\catcode `\$12\catcode
  `\&12\catcode `\#12\catcode `\^12\catcode `\_12\catcode `\%12\relax}%
\providecommand \@@startlink[1]{}%
\providecommand \@@endlink[0]{}%
\providecommand \url  [0]{\begingroup\@sanitize@url \@url }%
\providecommand \@url [1]{\endgroup\@href {#1}{\urlprefix }}%
\providecommand \urlprefix  [0]{URL }%
\providecommand \Eprint [0]{\href }%
\providecommand \doibase [0]{http://dx.doi.org/}%
\providecommand \selectlanguage [0]{\@gobble}%
\providecommand \bibinfo  [0]{\@secondoftwo}%
\providecommand \bibfield  [0]{\@secondoftwo}%
\providecommand \translation [1]{[#1]}%
\providecommand \BibitemOpen [0]{}%
\providecommand \bibitemStop [0]{}%
\providecommand \bibitemNoStop [0]{.\EOS\space}%
\providecommand \EOS [0]{\spacefactor3000\relax}%
\providecommand \BibitemShut  [1]{\csname bibitem#1\endcsname}%
\let\auto@bib@innerbib\@empty
%</preamble>
\bibitem [{\citenamefont {McDonnell}\ and\ \citenamefont
  {Ward}(2011)}]{McDonell2011}%
  \BibitemOpen
  \bibfield  {author} {\bibinfo {author} {\bibfnamefont {M.~D.}\ \bibnamefont
  {McDonnell}}\ and\ \bibinfo {author} {\bibfnamefont {L.~M.}\ \bibnamefont
  {Ward}},\ }\href@noop {} {\bibfield  {journal} {\bibinfo  {journal} {Nat Rev
  Neurosci}\ }\textbf {\bibinfo {volume} {12}},\ \bibinfo {pages} {415}
  (\bibinfo {year} {2011})}\BibitemShut {NoStop}%
\bibitem [{\citenamefont {Lindner}\ \emph {et~al.}(2004)\citenamefont
  {Lindner}, \citenamefont {Garcia-Ojalvo}, \citenamefont {Neiman},\ and\
  \citenamefont {Schimansky-Geier}}]{lindner04}%
  \BibitemOpen
  \bibfield  {author} {\bibinfo {author} {\bibfnamefont {B.}~\bibnamefont
  {Lindner}}, \bibinfo {author} {\bibfnamefont {J.}~\bibnamefont
  {Garcia-Ojalvo}}, \bibinfo {author} {\bibfnamefont {A.}~\bibnamefont
  {Neiman}}, \ and\ \bibinfo {author} {\bibfnamefont {L.}~\bibnamefont
  {Schimansky-Geier}},\ }\href@noop {} {\bibfield  {journal} {\bibinfo
  {journal} {Phys. Report}\ }\textbf {\bibinfo {volume} {392}},\ \bibinfo
  {pages} {321} (\bibinfo {year} {2004})}\BibitemShut {NoStop}%
\bibitem [{\citenamefont {Collins}\ \emph {et~al.}(1996)\citenamefont
  {Collins}, \citenamefont {Imhoff},\ and\ \citenamefont
  {Grigg}}]{collins1996}%
  \BibitemOpen
  \bibfield  {author} {\bibinfo {author} {\bibfnamefont {J.~J.}\ \bibnamefont
  {Collins}}, \bibinfo {author} {\bibfnamefont {T.~T.}\ \bibnamefont {Imhoff}},
  \ and\ \bibinfo {author} {\bibfnamefont {P.}~\bibnamefont {Grigg}},\
  }\href@noop {} {\bibfield  {journal} {\bibinfo  {journal} {Nature}\ }\textbf
  {\bibinfo {volume} {383}},\ \bibinfo {pages} {770} (\bibinfo {year}
  {1996})}\BibitemShut {NoStop}%
\bibitem [{\citenamefont {Bezrukov}\ and\ \citenamefont
  {Vodyanoy}(1995)}]{bezrukov95}%
  \BibitemOpen
  \bibfield  {author} {\bibinfo {author} {\bibfnamefont {S.~M.}\ \bibnamefont
  {Bezrukov}}\ and\ \bibinfo {author} {\bibfnamefont {I.}~\bibnamefont
  {Vodyanoy}},\ }\href@noop {} {\bibfield  {journal} {\bibinfo  {journal}
  {Nature}\ }\textbf {\bibinfo {volume} {378 (6555)}},\ \bibinfo {pages} {362}
  (\bibinfo {year} {1995})}\BibitemShut {NoStop}%
\bibitem [{\citenamefont {Ozer}\ \emph {et~al.}(2009)\citenamefont {Ozer},
  \citenamefont {Perc},\ and\ \citenamefont {Uzuntarla}}]{Ozer2009964}%
  \BibitemOpen
  \bibfield  {author} {\bibinfo {author} {\bibfnamefont {M.}~\bibnamefont
  {Ozer}}, \bibinfo {author} {\bibfnamefont {M.}~\bibnamefont {Perc}}, \ and\
  \bibinfo {author} {\bibfnamefont {M.}~\bibnamefont {Uzuntarla}},\ }\href
  {\doibase http://dx.doi.org/10.1016/j.physleta.2009.01.034} {\bibfield
  {journal} {\bibinfo  {journal} {Physics Letters A}\ }\textbf {\bibinfo
  {volume} {373}},\ \bibinfo {pages} {964 } (\bibinfo {year}
  {2009})}\BibitemShut {NoStop}%
\bibitem [{\citenamefont {Benzi}\ \emph {et~al.}(1981)\citenamefont {Benzi},
  \citenamefont {Sutera},\ and\ \citenamefont {Vulpiani}}]{benzi1981}%
  \BibitemOpen
  \bibfield  {author} {\bibinfo {author} {\bibfnamefont {R.}~\bibnamefont
  {Benzi}}, \bibinfo {author} {\bibfnamefont {A.}~\bibnamefont {Sutera}}, \
  and\ \bibinfo {author} {\bibfnamefont {A.}~\bibnamefont {Vulpiani}},\
  }\href@noop {} {\bibfield  {journal} {\bibinfo  {journal} {J. Phys. A: Math.
  Gen.}\ }\textbf {\bibinfo {volume} {14}},\ \bibinfo {pages} {L453} (\bibinfo
  {year} {1981})}\BibitemShut {NoStop}%
\bibitem [{\citenamefont {Collins}\ \emph {et~al.}(1995)\citenamefont
  {Collins}, \citenamefont {Chow},\ and\ \citenamefont
  {Imhoff}}]{collins_stochastic_1995}%
  \BibitemOpen
  \bibfield  {author} {\bibinfo {author} {\bibfnamefont {J.~J.}\ \bibnamefont
  {Collins}}, \bibinfo {author} {\bibfnamefont {C.}~\bibnamefont {Chow}}, \
  and\ \bibinfo {author} {\bibfnamefont {T.~T.}\ \bibnamefont {Imhoff}},\
  }\href@noop {} {\bibfield  {journal} {\bibinfo  {journal} {Nature}\ }\textbf
  {\bibinfo {volume} {376}},\ \bibinfo {pages} {236} (\bibinfo {year}
  {1995})}\BibitemShut {NoStop}%
\bibitem [{\citenamefont {Gammaitoni}\ \emph {et~al.}(1998)\citenamefont
  {Gammaitoni}, \citenamefont {Hanggi}, \citenamefont {Jung},\ and\
  \citenamefont {Marchesoni}}]{gammaitoni98}%
  \BibitemOpen
  \bibfield  {author} {\bibinfo {author} {\bibfnamefont {L.}~\bibnamefont
  {Gammaitoni}}, \bibinfo {author} {\bibfnamefont {P.}~\bibnamefont {Hanggi}},
  \bibinfo {author} {\bibfnamefont {P.}~\bibnamefont {Jung}}, \ and\ \bibinfo
  {author} {\bibfnamefont {F.}~\bibnamefont {Marchesoni}},\ }\href@noop {}
  {\bibfield  {journal} {\bibinfo  {journal} {Rev. Mod. Phys.}\ }\textbf
  {\bibinfo {volume} {70}},\ \bibinfo {pages} {223} (\bibinfo {year}
  {1998})}\BibitemShut {NoStop}%
\bibitem [{\citenamefont {Manjarrez}\ \emph {et~al.}(2002)\citenamefont
  {Manjarrez}, \citenamefont {Diez-Martinez}, \citenamefont {Mendez},\ and\
  \citenamefont {Flores}}]{manjarrez02}%
  \BibitemOpen
  \bibfield  {author} {\bibinfo {author} {\bibfnamefont {E.}~\bibnamefont
  {Manjarrez}}, \bibinfo {author} {\bibfnamefont {O.}~\bibnamefont
  {Diez-Martinez}}, \bibinfo {author} {\bibfnamefont {I.}~\bibnamefont
  {Mendez}}, \ and\ \bibinfo {author} {\bibfnamefont {A.}~\bibnamefont
  {Flores}},\ }\href@noop {} {\bibfield  {journal} {\bibinfo  {journal}
  {Neurosci. Lett.}\ }\textbf {\bibinfo {volume} {324 (3)}},\ \bibinfo {pages}
  {213} (\bibinfo {year} {2002})}\BibitemShut {NoStop}%
\bibitem [{\citenamefont {Tuckwell}\ \emph {et~al.}(2009)\citenamefont
  {Tuckwell}, \citenamefont {Jost},\ and\ \citenamefont
  {Gutkin}}]{TuckwellPhysRevE09}%
  \BibitemOpen
  \bibfield  {author} {\bibinfo {author} {\bibfnamefont {H.~C.}\ \bibnamefont
  {Tuckwell}}, \bibinfo {author} {\bibfnamefont {J.}~\bibnamefont {Jost}}, \
  and\ \bibinfo {author} {\bibfnamefont {B.~S.}\ \bibnamefont {Gutkin}},\
  }\href@noop {} {\bibfield  {journal} {\bibinfo  {journal} {Phys. Rev. E}\
  }\textbf {\bibinfo {volume} {80}},\ \bibinfo {pages} {031907} (\bibinfo
  {year} {2009})}\BibitemShut {NoStop}%
\bibitem [{\citenamefont {Guo}(2011)}]{Guo11}%
  \BibitemOpen
  \bibfield  {author} {\bibinfo {author} {\bibfnamefont {D.}~\bibnamefont
  {Guo}},\ }\href@noop {} {\bibfield  {journal} {\bibinfo  {journal} {Cogn.
  Neurodyn.}\ }\textbf {\bibinfo {volume} {5}},\ \bibinfo {pages} {293}
  (\bibinfo {year} {2011})}\BibitemShut {NoStop}%
\bibitem [{\citenamefont {Gutkin}\ \emph {et~al.}(2008)\citenamefont {Gutkin},
  \citenamefont {Jost},\ and\ \citenamefont {Tuckwell}}]{ISRtwoneurons}%
  \BibitemOpen
  \bibfield  {author} {\bibinfo {author} {\bibfnamefont {B.~S.}\ \bibnamefont
  {Gutkin}}, \bibinfo {author} {\bibfnamefont {J.}~\bibnamefont {Jost}}, \ and\
  \bibinfo {author} {\bibfnamefont {H.~C.}\ \bibnamefont {Tuckwell}},\
  }\href@noop {} {\bibfield  {journal} {\bibinfo  {journal} {EPL (Europhysics
  Letters)}\ }\textbf {\bibinfo {volume} {81}},\ \bibinfo {pages} {20005}
  (\bibinfo {year} {2008})}\BibitemShut {NoStop}%
\bibitem [{\citenamefont {Uzuntarla}(2013)}]{Uzuntarlasolo13}%
  \BibitemOpen
  \bibfield  {author} {\bibinfo {author} {\bibfnamefont {M.}~\bibnamefont
  {Uzuntarla}},\ }\href@noop {} {\bibfield  {journal} {\bibinfo  {journal}
  {Phys. Lett. A}\ }\textbf {\bibinfo {volume} {377}},\ \bibinfo {pages} {2585}
  (\bibinfo {year} {2013})}\BibitemShut {NoStop}%
\bibitem [{\citenamefont {Uzuntarla}\ \emph {et~al.}(2013)\citenamefont
  {Uzuntarla}, \citenamefont {Cressman}, \citenamefont {Ozer},\ and\
  \citenamefont {Barreto}}]{uzuntarla13}%
  \BibitemOpen
  \bibfield  {author} {\bibinfo {author} {\bibfnamefont {M.}~\bibnamefont
  {Uzuntarla}}, \bibinfo {author} {\bibfnamefont {J.~R.}\ \bibnamefont
  {Cressman}}, \bibinfo {author} {\bibfnamefont {M.}~\bibnamefont {Ozer}}, \
  and\ \bibinfo {author} {\bibfnamefont {E.}~\bibnamefont {Barreto}},\
  }\href@noop {} {\bibfield  {journal} {\bibinfo  {journal} {Phys. Rev. E}\
  }\textbf {\bibinfo {volume} {88}},\ \bibinfo {pages} {042712} (\bibinfo
  {year} {2013})}\BibitemShut {NoStop}%
\bibitem [{\citenamefont {Schmerl}\ and\ \citenamefont
  {McDonnell}(2013)}]{schmerl13}%
  \BibitemOpen
  \bibfield  {author} {\bibinfo {author} {\bibfnamefont {B.~A.}\ \bibnamefont
  {Schmerl}}\ and\ \bibinfo {author} {\bibfnamefont {M.~D.}\ \bibnamefont
  {McDonnell}},\ }\href@noop {} {\bibfield  {journal} {\bibinfo  {journal}
  {Phys. Rev. E}\ }\textbf {\bibinfo {volume} {88}},\ \bibinfo {pages} {052722}
  (\bibinfo {year} {2013})}\BibitemShut {NoStop}%
\bibitem [{\citenamefont {Paydarfar}\ \emph {et~al.}(2006)\citenamefont
  {Paydarfar}, \citenamefont {Forger},\ and\ \citenamefont
  {Clay}}]{paydarfar_noisy_2006}%
  \BibitemOpen
  \bibfield  {author} {\bibinfo {author} {\bibfnamefont {D.}~\bibnamefont
  {Paydarfar}}, \bibinfo {author} {\bibfnamefont {D.~B.}\ \bibnamefont
  {Forger}}, \ and\ \bibinfo {author} {\bibfnamefont {J.~R.}\ \bibnamefont
  {Clay}},\ }\href@noop {} {\bibfield  {journal} {\bibinfo  {journal} {Journal
  of Neurophysiology}\ }\textbf {\bibinfo {volume} {96}},\ \bibinfo {pages}
  {3338} (\bibinfo {year} {2006})}\BibitemShut {NoStop}%
\bibitem [{\citenamefont {Tuckwell}\ and\ \citenamefont
  {Jost}(2011)}]{tuckwell_effects_2011}%
  \BibitemOpen
  \bibfield  {author} {\bibinfo {author} {\bibfnamefont {H.~C.}\ \bibnamefont
  {Tuckwell}}\ and\ \bibinfo {author} {\bibfnamefont {J.}~\bibnamefont
  {Jost}},\ }\href@noop {} {\bibfield  {journal} {\bibinfo  {journal} {J Comput
  Neurosci}\ }\textbf {\bibinfo {volume} {30}},\ \bibinfo {pages} {361}
  (\bibinfo {year} {2011})}\BibitemShut {NoStop}%
\bibitem [{\citenamefont {Stein}\ \emph {et~al.}(2005)\citenamefont {Stein},
  \citenamefont {Gossen},\ and\ \citenamefont {Jones}}]{Stein2005}%
  \BibitemOpen
  \bibfield  {author} {\bibinfo {author} {\bibfnamefont {R.~B.}\ \bibnamefont
  {Stein}}, \bibinfo {author} {\bibfnamefont {E.~R.}\ \bibnamefont {Gossen}}, \
  and\ \bibinfo {author} {\bibfnamefont {K.~E.}\ \bibnamefont {Jones}},\ }\href
  {http://dx.doi.org/10.1038/nrn1668} {\bibfield  {journal} {\bibinfo
  {journal} {Nat Rev Neurosci}\ }\textbf {\bibinfo {volume} {6}},\ \bibinfo
  {pages} {389} (\bibinfo {year} {2005})}\BibitemShut {NoStop}%
\bibitem [{\citenamefont {Lin}\ \emph {et~al.}(2015)\citenamefont {Lin},
  \citenamefont {Okun}, \citenamefont {Carandini},\ and\ \citenamefont
  {Harris}}]{Lin2015644}%
  \BibitemOpen
  \bibfield  {author} {\bibinfo {author} {\bibfnamefont {I.-C.}\ \bibnamefont
  {Lin}}, \bibinfo {author} {\bibfnamefont {M.}~\bibnamefont {Okun}}, \bibinfo
  {author} {\bibfnamefont {M.}~\bibnamefont {Carandini}}, \ and\ \bibinfo
  {author} {\bibfnamefont {K.~D.}\ \bibnamefont {Harris}},\ }\href {\doibase
  http://dx.doi.org/10.1016/j.neuron.2015.06.035} {\bibfield  {journal}
  {\bibinfo  {journal} {Neuron}\ }\textbf {\bibinfo {volume} {87}},\ \bibinfo
  {pages} {644 } (\bibinfo {year} {2015})}\BibitemShut {NoStop}%
\bibitem [{\citenamefont {Franks}\ \emph {et~al.}(2003)\citenamefont {Franks},
  \citenamefont {Stevens},\ and\ \citenamefont {Sejnowski}}]{fssjn03}%
  \BibitemOpen
  \bibfield  {author} {\bibinfo {author} {\bibfnamefont {K.~M.}\ \bibnamefont
  {Franks}}, \bibinfo {author} {\bibfnamefont {C.~F.}\ \bibnamefont {Stevens}},
  \ and\ \bibinfo {author} {\bibfnamefont {T.~J.}\ \bibnamefont {Sejnowski}},\
  }\href@noop {} {\bibfield  {journal} {\bibinfo  {journal} {J. Neurosci.}\
  }\textbf {\bibinfo {volume} {23}},\ \bibinfo {pages} {3186} (\bibinfo {year}
  {2003})}\BibitemShut {NoStop}%
\bibitem [{\citenamefont {Zucker}\ and\ \citenamefont
  {Regehr}(2002)}]{zuckerARP02}%
  \BibitemOpen
  \bibfield  {author} {\bibinfo {author} {\bibfnamefont {R.~S.}\ \bibnamefont
  {Zucker}}\ and\ \bibinfo {author} {\bibfnamefont {W.~G.}\ \bibnamefont
  {Regehr}},\ }\href@noop {} {\bibfield  {journal} {\bibinfo  {journal} {Annu.
  Rev. Physiol.}\ }\textbf {\bibinfo {volume} {64}},\ \bibinfo {pages} {355}
  (\bibinfo {year} {2002})}\BibitemShut {NoStop}%
\bibitem [{\citenamefont {Dittman}\ \emph {et~al.}(2000)\citenamefont
  {Dittman}, \citenamefont {Kreitzer},\ and\ \citenamefont
  {Regehr}}]{dittmanJNEUROSCI}%
  \BibitemOpen
  \bibfield  {author} {\bibinfo {author} {\bibfnamefont {J.~S.}\ \bibnamefont
  {Dittman}}, \bibinfo {author} {\bibfnamefont {A.~C.}\ \bibnamefont
  {Kreitzer}}, \ and\ \bibinfo {author} {\bibfnamefont {W.~G.}\ \bibnamefont
  {Regehr}},\ }\href@noop {} {\bibfield  {journal} {\bibinfo  {journal} {J.
  Neurosci.}\ }\textbf {\bibinfo {volume} {20}},\ \bibinfo {pages} {1374}
  (\bibinfo {year} {2000})}\BibitemShut {NoStop}%
\bibitem [{\citenamefont {Cho}\ \emph {et~al.}(2011)\citenamefont {Cho},
  \citenamefont {Li},\ and\ \citenamefont {von Gersdorff}}]{Cho13042011}%
  \BibitemOpen
  \bibfield  {author} {\bibinfo {author} {\bibfnamefont {S.}~\bibnamefont
  {Cho}}, \bibinfo {author} {\bibfnamefont {G.-L.}\ \bibnamefont {Li}}, \ and\
  \bibinfo {author} {\bibfnamefont {H.}~\bibnamefont {von Gersdorff}},\ }\href
  {\doibase 10.1523/JNEUROSCI.5453-10.2011} {\bibfield  {journal} {\bibinfo
  {journal} {The Journal of Neuroscience}\ }\textbf {\bibinfo {volume} {31}},\
  \bibinfo {pages} {5682} (\bibinfo {year} {2011})}\BibitemShut {NoStop}%
\bibitem [{\citenamefont {de~la Rocha}\ and\ \citenamefont
  {Parga}(2005)}]{delarocha05}%
  \BibitemOpen
  \bibfield  {author} {\bibinfo {author} {\bibfnamefont {J.}~\bibnamefont
  {de~la Rocha}}\ and\ \bibinfo {author} {\bibfnamefont {N.}~\bibnamefont
  {Parga}},\ }\href@noop {} {\bibfield  {journal} {\bibinfo  {journal} {J.
  Neurosci.}\ }\textbf {\bibinfo {volume} {25}},\ \bibinfo {pages} {8416}
  (\bibinfo {year} {2005})}\BibitemShut {NoStop}%
\bibitem [{\citenamefont {Bertram}\ \emph {et~al.}(1996)\citenamefont
  {Bertram}, \citenamefont {Sherman},\ and\ \citenamefont
  {Stanley}}]{bertramJNEURO}%
  \BibitemOpen
  \bibfield  {author} {\bibinfo {author} {\bibfnamefont {R.}~\bibnamefont
  {Bertram}}, \bibinfo {author} {\bibfnamefont {A.}~\bibnamefont {Sherman}}, \
  and\ \bibinfo {author} {\bibfnamefont {E.~F.}\ \bibnamefont {Stanley}},\
  }\href@noop {} {\bibfield  {journal} {\bibinfo  {journal} {J. Neurophysiol.}\
  }\textbf {\bibinfo {volume} {75}},\ \bibinfo {pages} {1919} (\bibinfo {year}
  {1996})}\BibitemShut {NoStop}%
\bibitem [{\citenamefont {Dobrunz}\ and\ \citenamefont
  {Stevens}(1997)}]{synapticnoise}%
  \BibitemOpen
  \bibfield  {author} {\bibinfo {author} {\bibfnamefont {L.~E.}\ \bibnamefont
  {Dobrunz}}\ and\ \bibinfo {author} {\bibfnamefont {C.~F.}\ \bibnamefont
  {Stevens}},\ }\href@noop {} {\bibfield  {journal} {\bibinfo  {journal}
  {Neuron}\ }\textbf {\bibinfo {volume} {18}},\ \bibinfo {pages} {995}
  (\bibinfo {year} {1997})}\BibitemShut {NoStop}%
\bibitem [{\citenamefont {Tsodyks}\ \emph {et~al.}(1998)\citenamefont
  {Tsodyks}, \citenamefont {Pawelzik},\ and\ \citenamefont
  {Markram}}]{tsodyksNC}%
  \BibitemOpen
  \bibfield  {author} {\bibinfo {author} {\bibfnamefont {M.~V.}\ \bibnamefont
  {Tsodyks}}, \bibinfo {author} {\bibfnamefont {K.}~\bibnamefont {Pawelzik}}, \
  and\ \bibinfo {author} {\bibfnamefont {H.}~\bibnamefont {Markram}},\
  }\href@noop {} {\bibfield  {journal} {\bibinfo  {journal} {Neural Comput.}\
  }\textbf {\bibinfo {volume} {10}},\ \bibinfo {pages} {821} (\bibinfo {year}
  {1998})}\BibitemShut {NoStop}%
\bibitem [{\citenamefont {Senn}\ \emph {et~al.}(1998)\citenamefont {Senn},
  \citenamefont {Segev},\ and\ \citenamefont {Tsodyks}}]{Senn98}%
  \BibitemOpen
  \bibfield  {author} {\bibinfo {author} {\bibfnamefont {W.}~\bibnamefont
  {Senn}}, \bibinfo {author} {\bibfnamefont {I.}~\bibnamefont {Segev}}, \ and\
  \bibinfo {author} {\bibfnamefont {M.}~\bibnamefont {Tsodyks}},\ }\href@noop
  {} {\bibfield  {journal} {\bibinfo  {journal} {Neural Comput.}\ }\textbf
  {\bibinfo {volume} {10}},\ \bibinfo {pages} {815} (\bibinfo {year}
  {1998})}\BibitemShut {NoStop}%
\bibitem [{\citenamefont {Dror}\ and\ \citenamefont
  {Tsodyks}(2000)}]{Dror2000}%
  \BibitemOpen
  \bibfield  {author} {\bibinfo {author} {\bibfnamefont {G.}~\bibnamefont
  {Dror}}\ and\ \bibinfo {author} {\bibfnamefont {M.}~\bibnamefont {Tsodyks}},\
  }\href@noop {} {\bibfield  {journal} {\bibinfo  {journal} {Neurocomputing}\
  }\textbf {\bibinfo {volume} {32-33}},\ \bibinfo {pages} {365} (\bibinfo
  {year} {2000})}\BibitemShut {NoStop}%
\bibitem [{\citenamefont {Tsodyks}\ \emph {et~al.}(2000)\citenamefont
  {Tsodyks}, \citenamefont {Uziel},\ and\ \citenamefont
  {Markram}}]{tsodyksjn00}%
  \BibitemOpen
  \bibfield  {author} {\bibinfo {author} {\bibfnamefont {M.}~\bibnamefont
  {Tsodyks}}, \bibinfo {author} {\bibfnamefont {A.}~\bibnamefont {Uziel}}, \
  and\ \bibinfo {author} {\bibfnamefont {H.}~\bibnamefont {Markram}},\
  }\href@noop {} {\bibfield  {journal} {\bibinfo  {journal} {J. Neurosci.}\
  }\textbf {\bibinfo {volume} {20}},\ \bibinfo {pages} {RC50} (\bibinfo {year}
  {2000})}\BibitemShut {NoStop}%
\bibitem [{\citenamefont {Fuhrmann}\ \emph {et~al.}(2001)\citenamefont
  {Fuhrmann}, \citenamefont {Segev}, \citenamefont {Markram},\ and\
  \citenamefont {Tsodyks}}]{tsodyksCODING}%
  \BibitemOpen
  \bibfield  {author} {\bibinfo {author} {\bibfnamefont {G.}~\bibnamefont
  {Fuhrmann}}, \bibinfo {author} {\bibfnamefont {I.}~\bibnamefont {Segev}},
  \bibinfo {author} {\bibfnamefont {H.}~\bibnamefont {Markram}}, \ and\
  \bibinfo {author} {\bibfnamefont {M.}~\bibnamefont {Tsodyks}},\ }\href@noop
  {} {\bibfield  {journal} {\bibinfo  {journal} {J. Neurophysiol.}\ }\textbf
  {\bibinfo {volume} {87}},\ \bibinfo {pages} {140} (\bibinfo {year}
  {2001})}\BibitemShut {NoStop}%
\bibitem [{\citenamefont {Pantic}\ \emph {et~al.}(2002)\citenamefont {Pantic},
  \citenamefont {Torres}, \citenamefont {Kappen},\ and\ \citenamefont
  {Gielen}}]{torresNC}%
  \BibitemOpen
  \bibfield  {author} {\bibinfo {author} {\bibfnamefont {L.}~\bibnamefont
  {Pantic}}, \bibinfo {author} {\bibfnamefont {J.~J.}\ \bibnamefont {Torres}},
  \bibinfo {author} {\bibfnamefont {H.~J.}\ \bibnamefont {Kappen}}, \ and\
  \bibinfo {author} {\bibfnamefont {S.~C. A.~M.}\ \bibnamefont {Gielen}},\
  }\href@noop {} {\bibfield  {journal} {\bibinfo  {journal} {Neural Comput.}\
  }\textbf {\bibinfo {volume} {14}},\ \bibinfo {pages} {2903} (\bibinfo {year}
  {2002})}\BibitemShut {NoStop}%
\bibitem [{\citenamefont {Holcman}\ and\ \citenamefont
  {Tsodyks}(2006)}]{tsodyks06}%
  \BibitemOpen
  \bibfield  {author} {\bibinfo {author} {\bibfnamefont {D.}~\bibnamefont
  {Holcman}}\ and\ \bibinfo {author} {\bibfnamefont {M.}~\bibnamefont
  {Tsodyks}},\ }\href@noop {} {\bibfield  {journal} {\bibinfo  {journal} {PLoS
  Comput. Biol.}\ }\textbf {\bibinfo {volume} {2}},\ \bibinfo {pages} {174}
  (\bibinfo {year} {2006})}\BibitemShut {NoStop}%
\bibitem [{\citenamefont {Barak}\ and\ \citenamefont
  {Tsodyks}(2007)}]{barak2007}%
  \BibitemOpen
  \bibfield  {author} {\bibinfo {author} {\bibfnamefont {O.}~\bibnamefont
  {Barak}}\ and\ \bibinfo {author} {\bibfnamefont {M.}~\bibnamefont
  {Tsodyks}},\ }\href@noop {} {\bibfield  {journal} {\bibinfo  {journal} {PLoS
  Comput. Biol}\ }\textbf {\bibinfo {volume} {3}},\ \bibinfo {pages} {e35}
  (\bibinfo {year} {2007})}\BibitemShut {NoStop}%
\bibitem [{\citenamefont {Torres}\ \emph {et~al.}(2008)\citenamefont {Torres},
  \citenamefont {Cortes}, \citenamefont {Marro},\ and\ \citenamefont
  {Kappen}}]{torresNC2007}%
  \BibitemOpen
  \bibfield  {author} {\bibinfo {author} {\bibfnamefont {J.~J.}\ \bibnamefont
  {Torres}}, \bibinfo {author} {\bibfnamefont {J.~M.}\ \bibnamefont {Cortes}},
  \bibinfo {author} {\bibfnamefont {J.}~\bibnamefont {Marro}}, \ and\ \bibinfo
  {author} {\bibfnamefont {H.~J.}\ \bibnamefont {Kappen}},\ }\href@noop {}
  {\bibfield  {journal} {\bibinfo  {journal} {Neural Comput.}\ }\textbf
  {\bibinfo {volume} {19}},\ \bibinfo {pages} {2739} (\bibinfo {year}
  {2008})}\BibitemShut {NoStop}%
\bibitem [{\citenamefont {Mejias}\ and\ \citenamefont
  {Torres}(2008)}]{mejiasCD08}%
  \BibitemOpen
  \bibfield  {author} {\bibinfo {author} {\bibfnamefont {J.~F.}\ \bibnamefont
  {Mejias}}\ and\ \bibinfo {author} {\bibfnamefont {J.~J.}\ \bibnamefont
  {Torres}},\ }\href@noop {} {\bibfield  {journal} {\bibinfo  {journal} {J.
  Comput. Neurosci.}\ }\textbf {\bibinfo {volume} {24}},\ \bibinfo {pages}
  {222} (\bibinfo {year} {2008})}\BibitemShut {NoStop}%
\bibitem [{\citenamefont {Mejias}\ and\ \citenamefont
  {Torres}(2011)}]{mejias2011emergence}%
  \BibitemOpen
  \bibfield  {author} {\bibinfo {author} {\bibfnamefont {J.~F.}\ \bibnamefont
  {Mejias}}\ and\ \bibinfo {author} {\bibfnamefont {J.~J.}\ \bibnamefont
  {Torres}},\ }\href@noop {} {\bibfield  {journal} {\bibinfo  {journal} {PLoS
  ONE}\ }\textbf {\bibinfo {volume} {6}},\ \bibinfo {pages} {e17255} (\bibinfo
  {year} {2011})}\BibitemShut {NoStop}%
\bibitem [{\citenamefont {Torres}\ and\ \citenamefont
  {Kappen}(2013)}]{TorresKappen2013}%
  \BibitemOpen
  \bibfield  {author} {\bibinfo {author} {\bibfnamefont {J.~J.}\ \bibnamefont
  {Torres}}\ and\ \bibinfo {author} {\bibfnamefont {J.~H.}\ \bibnamefont
  {Kappen}},\ }\href@noop {} {\bibfield  {journal} {\bibinfo  {journal} {Front.
  Comput. Neurosci.}\ }\textbf {\bibinfo {volume} {7}},\ \bibinfo {pages} {30}
  (\bibinfo {year} {2013})}\BibitemShut {NoStop}%
\bibitem [{\citenamefont {Torres}\ \emph {et~al.}(2002)\citenamefont {Torres},
  \citenamefont {Pantic},\ and\ \citenamefont {Kappen}}]{torresCAPACITY}%
  \BibitemOpen
  \bibfield  {author} {\bibinfo {author} {\bibfnamefont {J.~J.}\ \bibnamefont
  {Torres}}, \bibinfo {author} {\bibfnamefont {L.}~\bibnamefont {Pantic}}, \
  and\ \bibinfo {author} {\bibfnamefont {H.~J.}\ \bibnamefont {Kappen}},\
  }\href@noop {} {\bibfield  {journal} {\bibinfo  {journal} {Phys. Rev. E.}\
  }\textbf {\bibinfo {volume} {66}},\ \bibinfo {pages} {061910} (\bibinfo
  {year} {2002})}\BibitemShut {NoStop}%
\bibitem [{\citenamefont {MacLeod}\ \emph {et~al.}(2007)\citenamefont
  {MacLeod}, \citenamefont {Horiuchi},\ and\ \citenamefont
  {Carr}}]{MacLeod2863}%
  \BibitemOpen
  \bibfield  {author} {\bibinfo {author} {\bibfnamefont {K.~M.}\ \bibnamefont
  {MacLeod}}, \bibinfo {author} {\bibfnamefont {T.~K.}\ \bibnamefont
  {Horiuchi}}, \ and\ \bibinfo {author} {\bibfnamefont {C.~E.}\ \bibnamefont
  {Carr}},\ }\href@noop {} {\bibfield  {journal} {\bibinfo  {journal} {Journal
  of Neurophysiology}\ }\textbf {\bibinfo {volume} {97}},\ \bibinfo {pages}
  {2863} (\bibinfo {year} {2007})}\BibitemShut {NoStop}%
\bibitem [{\citenamefont {Fortune}\ and\ \citenamefont
  {Rose}(2002)}]{Fortune2002539}%
  \BibitemOpen
  \bibfield  {author} {\bibinfo {author} {\bibfnamefont {E.~S.}\ \bibnamefont
  {Fortune}}\ and\ \bibinfo {author} {\bibfnamefont {G.~J.}\ \bibnamefont
  {Rose}},\ }\href@noop {} {\bibfield  {journal} {\bibinfo  {journal} {Journal
  of Physiology-Paris}\ }\textbf {\bibinfo {volume} {96}},\ \bibinfo {pages}
  {539 } (\bibinfo {year} {2002})}\BibitemShut {NoStop}%
\bibitem [{\citenamefont {Mejias}\ and\ \citenamefont
  {Torres}(2009)}]{mejias09}%
  \BibitemOpen
  \bibfield  {author} {\bibinfo {author} {\bibfnamefont {J.~F.}\ \bibnamefont
  {Mejias}}\ and\ \bibinfo {author} {\bibfnamefont {J.~J.}\ \bibnamefont
  {Torres}},\ }\href@noop {} {\bibfield  {journal} {\bibinfo  {journal} {Neural
  Comput.}\ }\textbf {\bibinfo {volume} {21}},\ \bibinfo {pages} {851}
  (\bibinfo {year} {2009})}\BibitemShut {NoStop}%
\bibitem [{\citenamefont {Mejias}\ \emph {et~al.}(2010)\citenamefont {Mejias},
  \citenamefont {Kappen},\ and\ \citenamefont {Torres}}]{mejiasupdown10}%
  \BibitemOpen
  \bibfield  {author} {\bibinfo {author} {\bibfnamefont {J.~F.}\ \bibnamefont
  {Mejias}}, \bibinfo {author} {\bibfnamefont {H.~J.}\ \bibnamefont {Kappen}},
  \ and\ \bibinfo {author} {\bibfnamefont {J.~J.}\ \bibnamefont {Torres}},\
  }\href@noop {} {\bibfield  {journal} {\bibinfo  {journal} {PLoS ONE}\
  }\textbf {\bibinfo {volume} {5}},\ \bibinfo {pages} {e13651} (\bibinfo {year}
  {2010})}\BibitemShut {NoStop}%
\bibitem [{\citenamefont {Bourjaily}\ and\ \citenamefont
  {Miller}(2012)}]{Bourjaily513}%
  \BibitemOpen
  \bibfield  {author} {\bibinfo {author} {\bibfnamefont {M.~A.}\ \bibnamefont
  {Bourjaily}}\ and\ \bibinfo {author} {\bibfnamefont {P.}~\bibnamefont
  {Miller}},\ }\href@noop {} {\bibfield  {journal} {\bibinfo  {journal}
  {Journal of Neurophysiology}\ }\textbf {\bibinfo {volume} {108}},\ \bibinfo
  {pages} {513} (\bibinfo {year} {2012})}\BibitemShut {NoStop}%
\bibitem [{\citenamefont {Mejias}\ \emph {et~al.}(2012)\citenamefont {Mejias},
  \citenamefont {Hernandez-Gomez},\ and\ \citenamefont {Torres}}]{mejias12}%
  \BibitemOpen
  \bibfield  {author} {\bibinfo {author} {\bibfnamefont {J.~F.}\ \bibnamefont
  {Mejias}}, \bibinfo {author} {\bibfnamefont {B.}~\bibnamefont
  {Hernandez-Gomez}}, \ and\ \bibinfo {author} {\bibfnamefont {J.~J.}\
  \bibnamefont {Torres}},\ }\href@noop {} {\bibfield  {journal} {\bibinfo
  {journal} {EPL (Europhysics Letters)}\ }\textbf {\bibinfo {volume} {97}},\
  \bibinfo {pages} {48008} (\bibinfo {year} {2012})}\BibitemShut {NoStop}%
\bibitem [{\citenamefont {Hodgkin}\ and\ \citenamefont {Huxley}(1952)}]{HHb}%
  \BibitemOpen
  \bibfield  {author} {\bibinfo {author} {\bibfnamefont {A.~L.}\ \bibnamefont
  {Hodgkin}}\ and\ \bibinfo {author} {\bibfnamefont {A.~F.}\ \bibnamefont
  {Huxley}},\ }\href@noop {} {\bibfield  {journal} {\bibinfo  {journal} {J.
  Physiol.}\ }\textbf {\bibinfo {volume} {117}},\ \bibinfo {pages} {500}
  (\bibinfo {year} {1952})}\BibitemShut {NoStop}%
\bibitem [{\citenamefont {Pankratova}\ and\ \citenamefont
  {Polovinkin}(2005)}]{pankratova2005b}%
  \BibitemOpen
  \bibfield  {author} {\bibinfo {author} {\bibfnamefont {E.~V.}\ \bibnamefont
  {Pankratova}}\ and\ \bibinfo {author} {\bibfnamefont {A.~V.}\ \bibnamefont
  {Polovinkin}},\ }\href@noop {} {\bibfield  {journal} {\bibinfo  {journal}
  {Eur. Phys. J. B}\ }\textbf {\bibinfo {volume} {45}},\ \bibinfo {pages} {391}
  (\bibinfo {year} {2005})}\BibitemShut {NoStop}%
\bibitem [{\citenamefont {Tsodyks}\ and\ \citenamefont
  {Markram}(1997)}]{tsodyksPNAS}%
  \BibitemOpen
  \bibfield  {author} {\bibinfo {author} {\bibfnamefont {M.~V.}\ \bibnamefont
  {Tsodyks}}\ and\ \bibinfo {author} {\bibfnamefont {H.}~\bibnamefont
  {Markram}},\ }\href@noop {} {\bibfield  {journal} {\bibinfo  {journal} {Proc.
  Natl. Acad. Sci. USA}\ }\textbf {\bibinfo {volume} {94}},\ \bibinfo {pages}
  {719} (\bibinfo {year} {1997})}\BibitemShut {NoStop}%
\bibitem [{\citenamefont {Brunel}(2000)}]{brunel00}%
  \BibitemOpen
  \bibfield  {author} {\bibinfo {author} {\bibfnamefont {N.}~\bibnamefont
  {Brunel}},\ }\href@noop {} {\bibfield  {journal} {\bibinfo  {journal} {J.
  Comp. Neurosci.}\ }\textbf {\bibinfo {volume} {8}},\ \bibinfo {pages} {183}
  (\bibinfo {year} {2000})}\BibitemShut {NoStop}%
\bibitem [{\citenamefont {Braitenberg}\ and\ \citenamefont
  {Schuz}(1991)}]{Braitenberg91}%
  \BibitemOpen
  \bibfield  {author} {\bibinfo {author} {\bibfnamefont {V.}~\bibnamefont
  {Braitenberg}}\ and\ \bibinfo {author} {\bibfnamefont {A.}~\bibnamefont
  {Schuz}},\ }\href@noop {} {\emph {\bibinfo {title} {Anatomy of the cortex:
  statistics and geometry}}}\ (\bibinfo  {publisher} {Springer, Berlin},\
  \bibinfo {year} {1991})\BibitemShut {NoStop}%
\bibitem [{\citenamefont {Gutkin}\ \emph {et~al.}(2009)\citenamefont {Gutkin},
  \citenamefont {Jost},\ and\ \citenamefont
  {Tuckwell}}]{gutkin_inhibition_2009}%
  \BibitemOpen
  \bibfield  {author} {\bibinfo {author} {\bibfnamefont {B.~S.}\ \bibnamefont
  {Gutkin}}, \bibinfo {author} {\bibfnamefont {J.}~\bibnamefont {Jost}}, \ and\
  \bibinfo {author} {\bibfnamefont {H.~C.}\ \bibnamefont {Tuckwell}},\
  }\href@noop {} {\bibfield  {journal} {\bibinfo  {journal}
  {Naturwissenschaften}\ }\textbf {\bibinfo {volume} {96}},\ \bibinfo {pages}
  {1091} (\bibinfo {year} {2009})}\BibitemShut {NoStop}%
\bibitem [{Note1()}]{Note1}%
  \BibitemOpen
  \bibinfo {note} {The value of $\nu $ can be predicted based on how the
  parameter space from which the initial condition is selected is proportioned
  with respect to the basins of the stable limit cycle and equilibrium. See
  \cite {uzuntarla13}.}\BibitemShut {Stop}%
\bibitem [{\citenamefont {Fitzpatrick}\ \emph {et~al.}(2001)\citenamefont
  {Fitzpatrick}, \citenamefont {Akopian},\ and\ \citenamefont
  {Walsh}}]{Fitzpatrick2001}%
  \BibitemOpen
  \bibfield  {author} {\bibinfo {author} {\bibfnamefont {J.~S.}\ \bibnamefont
  {Fitzpatrick}}, \bibinfo {author} {\bibfnamefont {G.}~\bibnamefont
  {Akopian}}, \ and\ \bibinfo {author} {\bibfnamefont {J.~P.}\ \bibnamefont
  {Walsh}},\ }\href@noop {} {\bibfield  {journal} {\bibinfo  {journal} {J.
  Neurophysiol.}\ }\textbf {\bibinfo {volume} {85}},\ \bibinfo {pages} {2088}
  (\bibinfo {year} {2001})}\BibitemShut {NoStop}%
\bibitem [{\citenamefont {Ma}\ \emph {et~al.}(2012)\citenamefont {Ma},
  \citenamefont {Hu},\ and\ \citenamefont {Agmon}}]{Ma2012}%
  \BibitemOpen
  \bibfield  {author} {\bibinfo {author} {\bibfnamefont {Y.}~\bibnamefont
  {Ma}}, \bibinfo {author} {\bibfnamefont {H.}~\bibnamefont {Hu}}, \ and\
  \bibinfo {author} {\bibfnamefont {A.}~\bibnamefont {Agmon}},\ }\href@noop {}
  {\bibfield  {journal} {\bibinfo  {journal} {J. Neurosci.}\ }\textbf {\bibinfo
  {volume} {32}},\ \bibinfo {pages} {983} (\bibinfo {year} {2012})}\BibitemShut
  {NoStop}%
\bibitem [{\citenamefont {Flores}\ \emph {et~al.}(2015)\citenamefont {Flores},
  \citenamefont {Valdez}, \citenamefont {Huerta}, \citenamefont {Galarraga},\
  and\ \citenamefont {Bargas}}]{Flores2015}%
  \BibitemOpen
  \bibfield  {author} {\bibinfo {author} {\bibfnamefont {J.~B.}\ \bibnamefont
  {Flores}}, \bibinfo {author} {\bibfnamefont {M.~A.~H.}\ \bibnamefont
  {Valdez}}, \bibinfo {author} {\bibfnamefont {V.~G.~L.}\ \bibnamefont
  {Huerta}}, \bibinfo {author} {\bibfnamefont {E.}~\bibnamefont {Galarraga}}, \
  and\ \bibinfo {author} {\bibfnamefont {J.}~\bibnamefont {Bargas}},\
  }\href@noop {} {\bibfield  {journal} {\bibinfo  {journal} {Neural Plast.}\
  }\textbf {\bibinfo {volume} {2015}},\ \bibinfo {pages} {573543} (\bibinfo
  {year} {2015})}\BibitemShut {NoStop}%
\bibitem [{Note2()}]{Note2}%
  \BibitemOpen
  \bibinfo {note} {More precisely, the boundary is the closure of the stable
  manifold of the ULC.}\BibitemShut {Stop}%
\bibitem [{\citenamefont {Rowat}(2007)}]{Rowat2007}%
  \BibitemOpen
  \bibfield  {author} {\bibinfo {author} {\bibfnamefont {P.}~\bibnamefont
  {Rowat}},\ }\href {\doibase 10.1162/neco.2007.19.5.1215} {\bibfield
  {journal} {\bibinfo  {journal} {Neural Computation}\ }\textbf {\bibinfo
  {volume} {19}},\ \bibinfo {pages} {1215} (\bibinfo {year}
  {2007})}\BibitemShut {NoStop}%
\bibitem [{\citenamefont {Buchin}\ \emph {et~al.}(2016)\citenamefont {Buchin},
  \citenamefont {Rieubland}, \citenamefont {H{\"a}usser}, \citenamefont
  {Gutkin},\ and\ \citenamefont {Roth}}]{Buchin2016}%
  \BibitemOpen
  \bibfield  {author} {\bibinfo {author} {\bibfnamefont {A.}~\bibnamefont
  {Buchin}}, \bibinfo {author} {\bibfnamefont {S.}~\bibnamefont {Rieubland}},
  \bibinfo {author} {\bibfnamefont {M.}~\bibnamefont {H{\"a}usser}}, \bibinfo
  {author} {\bibfnamefont {B.~S.}\ \bibnamefont {Gutkin}}, \ and\ \bibinfo
  {author} {\bibfnamefont {A.}~\bibnamefont {Roth}},\ }\href {\doibase
  10.1371/journal.pcbi.1005000} {\bibfield  {journal} {\bibinfo  {journal}
  {PLoS Comput Biol}\ }\textbf {\bibinfo {volume} {12}},\ \bibinfo {pages} {1}
  (\bibinfo {year} {2016})}\BibitemShut {NoStop}%
\bibitem [{\citenamefont {Purves}\ \emph {et~al.}(2012)\citenamefont {Purves},
  \citenamefont {Augustine}, \citenamefont {Fitzpatrick}, \citenamefont {Hall},
  \citenamefont {LaMantia},\ and\ \citenamefont {White}}]{Purves}%
  \BibitemOpen
  \bibfield  {author} {\bibinfo {author} {\bibfnamefont {D.}~\bibnamefont
  {Purves}}, \bibinfo {author} {\bibfnamefont {G.~J.}\ \bibnamefont
  {Augustine}}, \bibinfo {author} {\bibfnamefont {D.}~\bibnamefont
  {Fitzpatrick}}, \bibinfo {author} {\bibfnamefont {W.~C.}\ \bibnamefont
  {Hall}}, \bibinfo {author} {\bibfnamefont {A.-S.}\ \bibnamefont {LaMantia}},
  \ and\ \bibinfo {author} {\bibfnamefont {L.~E.}\ \bibnamefont {White}},\
  }\href@noop {} {\emph {\bibinfo {title} {Neuroscience}}},\ \bibinfo {edition}
  {5th}\ ed.\ (\bibinfo  {publisher} {Sinauer Associates, Inc.},\ \bibinfo
  {year} {2012})\BibitemShut {NoStop}%
\bibitem [{\citenamefont {Nagy}\ \emph {et~al.}(2005)\citenamefont {Nagy},
  \citenamefont {Milosevic}, \citenamefont {Fasshauer}, \citenamefont
  {M\"{u}ller}, \citenamefont {de~Groot}, \citenamefont {Lang}, \citenamefont
  {Wilson},\ and\ \citenamefont {S{\o}rensen}}]{Nagy}%
  \BibitemOpen
  \bibfield  {author} {\bibinfo {author} {\bibfnamefont {G.}~\bibnamefont
  {Nagy}}, \bibinfo {author} {\bibfnamefont {I.}~\bibnamefont {Milosevic}},
  \bibinfo {author} {\bibfnamefont {D.}~\bibnamefont {Fasshauer}}, \bibinfo
  {author} {\bibfnamefont {E.~M.}\ \bibnamefont {M\"{u}ller}}, \bibinfo
  {author} {\bibfnamefont {B.~L.}\ \bibnamefont {de~Groot}}, \bibinfo {author}
  {\bibfnamefont {T.}~\bibnamefont {Lang}}, \bibinfo {author} {\bibfnamefont
  {M.~C.}\ \bibnamefont {Wilson}}, \ and\ \bibinfo {author} {\bibfnamefont
  {J.~B.}\ \bibnamefont {S{\o}rensen}},\ }\href@noop {} {\bibfield  {journal}
  {\bibinfo  {journal} {Molecular Biology of the Cell}\ }\textbf {\bibinfo
  {volume} {16}},\ \bibinfo {pages} {5675} (\bibinfo {year}
  {2005})}\BibitemShut {NoStop}%
\bibitem [{\citenamefont {Dipoppa}\ and\ \citenamefont
  {Gutkin}(2013)}]{Dipoppa2013}%
  \BibitemOpen
  \bibfield  {author} {\bibinfo {author} {\bibfnamefont {M.}~\bibnamefont
  {Dipoppa}}\ and\ \bibinfo {author} {\bibfnamefont {B.~S.}\ \bibnamefont
  {Gutkin}},\ }\href {\doibase 10.1073/pnas.1303270110} {\bibfield  {journal}
  {\bibinfo  {journal} {Proceedings of the National Academy of Sciences}\
  }\textbf {\bibinfo {volume} {110}},\ \bibinfo {pages} {12828} (\bibinfo
  {year} {2013})}\BibitemShut {NoStop}%
\bibitem [{\citenamefont {De~Pitt{\'a}}\ and\ \citenamefont
  {Brunel}(2016)}]{PittaBrunel2016}%
  \BibitemOpen
  \bibfield  {author} {\bibinfo {author} {\bibfnamefont {M.}~\bibnamefont
  {De~Pitt{\'a}}}\ and\ \bibinfo {author} {\bibfnamefont {N.}~\bibnamefont
  {Brunel}},\ }\href@noop {} {\bibfield  {journal} {\bibinfo  {journal} {Neural
  Plast.}\ }\textbf {\bibinfo {volume} {2016}},\ \bibinfo {pages} {7607924}
  (\bibinfo {year} {2016})}\BibitemShut {NoStop}%
\bibitem [{\citenamefont {De~Pitt{\'a}}\ \emph {et~al.}(2016)\citenamefont
  {De~Pitt{\'a}}, \citenamefont {Brunel},\ and\ \citenamefont
  {Volterra}}]{DePitta2016}%
  \BibitemOpen
  \bibfield  {author} {\bibinfo {author} {\bibfnamefont {M.}~\bibnamefont
  {De~Pitt{\'a}}}, \bibinfo {author} {\bibfnamefont {N.}~\bibnamefont
  {Brunel}}, \ and\ \bibinfo {author} {\bibfnamefont {A.}~\bibnamefont
  {Volterra}},\ }\href {\doibase
  http://dx.doi.org/10.1016/j.neuroscience.2015.04.001} {\bibfield  {journal}
  {\bibinfo  {journal} {Neuroscience}\ }\textbf {\bibinfo {volume} {323}},\
  \bibinfo {pages} {43 } (\bibinfo {year} {2016})}\BibitemShut {NoStop}%
\end{thebibliography}

%

\end{document}